\pdfoutput=1 
\documentclass[cernpreprint,english,USenglish]{na61doc}

\usepackage{lineno, blindtext}
\usepackage{amsmath}
\usepackage{enumerate}
\usepackage{placeins}
\usepackage{appendix}
\usepackage[LY1]{fontenc}
\usepackage[utf8]{inputenc}
\usepackage{chngcntr}
\usepackage{microtype}
\usepackage{lineno}
\usepackage{color}
\usepackage{epstopdf}
\usepackage{colortbl}
\definecolor{darkred}{rgb}{0.5,0,0}
\definecolor{darkblue}{rgb}{0,0,0.5}
\definecolor{firebrick}{rgb}{0.75,0.125,0.125}
\definecolor{darkgreen}{rgb}{0,0.5,0}
\graphicspath{{Figures/}}

\usepackage{graphicx}
\usepackage{booktabs}
\usepackage{tabularx}
\usepackage{dcolumn}
\usepackage{bm}
\usepackage{xspace} 
\usepackage{longtable}
\usepackage{caption} 
\usepackage{url}
\usepackage{lmodern} 

\usepackage[colorinlistoftodos]{todonotes}

\newcommand{\nova}{NO{$\nu$}A }
\newcommand{\minerva}{MINER{$\nu$}A }
\newcommand{\gev}{\mbox{GeV}\xspace}
\newcommand{\gevc}{\mbox{GeV/$c$}\xspace}

\newcommand{\mev}{\mbox{MeV}\xspace}
\newcommand{\mevc}{\mbox{MeV/$c$}\xspace}

\newcommand{\piBe}{\ensuremath{\pi^+ + \textup{Be at 60}\:\gevc}\xspace}
\newcommand{\piCnoE}{\ensuremath{\pi^+ + \textup{C}}\xspace}

\newcommand{\pip}{\ensuremath{\pi^+}\xspace}
\newcommand{\pim}{\ensuremath{\pi^-}\xspace}
\newcommand{\pipm}{\ensuremath{\pi^\pm}\xspace}

\newcommand{\kos}{\ensuremath{K^0_{\textrm{S}}}\xspace}

\newcommand{\p}{\ensuremath{p}\xspace}
\newcommand{\ap}{\ensuremath{\bar{p}}\xspace}
\newcommand{\lam}{\ensuremath{\Lambda}\xspace}
\newcommand{\alam}{\ensuremath{\bar{\Lambda}}\xspace}
\newcommand{\vo}{\ensuremath{V^0}\xspace}
\newcommand{\vos}{\ensuremath{V^0}s\xspace}

\newcommand{\GeantFour}{{\textsc{Geant4}}\xspace}

\newcommand{\GeantFourVersion}{{\textsc{Geant}\xspace4.10.7}\xspace}

\ShineTitle{Measurements of \kos, \lam and \alam production in 120 GeV/$c$ p + C interactions}

\PreprintIdNumber{CERN-EP-2022-YYY}

\ShineJournal{Phys. Rev. D}
\ShineAbstract{ 
This paper presents multiplicity measurements of \kos, \lam, and \alam produced in 120 \gevc proton-carbon interactions. The measurements were made using data collected at the NA61/SHINE experiment during two different periods. Decays of these neutral hadrons impact the measured \pip, \pim, \p and \ap multiplicities in the 120 \gevc proton-carbon reaction, which are crucial inputs for long-baseline neutrino experiment predictions of neutrino beam flux. The double-differential multiplicities presented here will be used to more precisely measure charged-hadron multiplicities in this reaction, and to re-weight neutral hadron production in neutrino beam Monte Carlo simulations.
}
\begin{document}

\maketitle


\newpage 
{\Large The \NASixtyOne Collaboration}
\bigskip


\noindent
H.~Adhikary$^{\,13}$,
K.K.~Allison$^{\,30}$,
N.~Amin$^{\,5}$,
E.V.~Andronov$^{\,25}$,
T.~Anti\'ci\'c$^{\,3}$,
I.-C.~Arsene$^{\,12}$,
Y.~Balkova$^{\,18}$,
M.~Baszczyk$^{\,17}$,
D.~Battaglia$^{\,29}$,
S.~Bhosale$^{\,14}$
% ...


\noindent
% ...
$^{3}$~Ru{\dj}er Bo\v{s}kovi\'c Institute, Zagreb, Croatia\\
% ...
$^{5}$~Karlsruhe Institute of Technology, Karlsruhe, Germany\\
% ...
$^{12}$~University of Oslo, Oslo, Norway\\
$^{13}$~Jan Kochanowski University in Kielce, Poland\\
$^{14}$~Institute of Nuclear Physics, Polish Academy of Sciences, Cracow, Poland\\
% ...
$^{17}$~AGH - University of Science and Technology, Cracow, Poland\\
$^{18}$~University of Silesia, Katowice, Poland\\
% ...
$^{25}$~St. Petersburg State University, St. Petersburg, Russia\\
% ...
$^{29}$~University of Notre Dame, Notre Dame , USA\\
$^{30}$~University of Colorado, Boulder, USA\\
% ...






\section{Introduction}

Measuring charged and neutral hadron production in the 120 \gevc proton-carbon interaction is crucial for predicting neutrino beam flux in current and future long-baseline neutrino oscillation experiments at Fermilab. This particular reaction is used by the NuMI facility at Fermilab to initiate the neutrino beam for the \nova experiment, and was used to produce the neutrino beam for the \minerva and MINOS experiments~\cite{numibeamline}. The future Long-Baseline Neutrino Facility (LBNF), which will provide the neutrino beam for the Deep Underground Neutrino Experiment (DUNE), will likely use the same primary reaction to create its beam~\cite{dune_physics}. 

A significant fraction of charged hadrons produced in 120 \gevc proton-carbon interactions originate from the decays of produced \kos, \lam, and \alam, which will be referred to throughout this manuscript as feed-down decays. In neutrino beam simulations, feed-down decay contribution to charged-hadron production is typically estimated using a \GeantFour-based simulation. However, the predicted fraction of charged particles produced via feed-down decay varies significantly depending on the \GeantFour physics list chosen (see Table~\ref{tab:feedDownFraction}). The need to rely on simulations to predict feed-down contribution can be eliminated if the \kos, \lam, and \alam multiplicities are measured directly.

\begin{table*}[htbp]
\centering
\begin{tabular}{cccccccc}
Hadron Species & FTFP\_BERT & QGSP\_BERT & FTF\_BIC \\
\hline
\pip  & 3.7\% & 5.4\% & 3.7\% \\
\pim  & 5.3\% & 7.5\% & 5.5\% \\
\p    & 5.9\% & 8.1\% & 5.0\% \\
\ap   &  23\% &  43\% &  23\% 

\end{tabular}
\caption[Feed Down Fractions from Physics Lists]{Fractions of charged hadrons produced by decay of \kos, \lam, or \alam as predicted by three commonly used \GeantFour physics lists.}
\label{tab:feedDownFraction}
\end{table*}

The NA61/SPS Heavy Ion and Neutrino Experiment (NA61/SHINE) is a fixed-target experiment located at the CERN Super Proton Synchrotron (SPS). NA61/SHINE makes dedicated hadron production measurements in reactions relevant to neutrino beam production. Hadron production measurements made at NA61/SHINE have been successfully used to improve neutrino flux estimates at existing long-baseline neutrino experiments~\cite{Abgrall:2011ae, Abgrall:2011ts, Abgrall:2013wda, Abgrall:2015hmv, Abgrall:2012pp, Abgrall:2016jif, Berns:2018tap, Acharya:2020jtn}. NA61/SHINE has published several papers specifically targeting hadron production in reactions relevant to Fermilab neutrino experiments~\cite{Aduszkiewicz:2019hhe,Aduszkiewicz:2018uts,Aduszkiewicz:2019xna}.

In 2016 and 2017, NA61/SHINE recorded two data sets measuring hadron production in 120 \gevc protons on a thin carbon target (3.1\% proton-nuclear interaction length, $\lambda$). Charged and neutral hadron analyses were performed on the recorded data sets, and the resulting multiplicities and uncertainties were combined where possible. Measurements presented in this publication will be used to reduce the uncertainties associated with weak neutral hadron decays in the charged-hadron analyses.

The measurements reported in this publication are performed using the main decay modes of each neutral hadron species:  $\kos \to \pip \; \pim$ (69.2\%), $\lam \to \p \; \pim$ (63.9\%) and $\alam \to \ap \; \pip$ (63.9\%)\cite{PDG}.

\section{Experimental Setup}\label{sec:Setup}

\begin{figure*}[t]
  \centering
  \includegraphics[width=\textwidth]{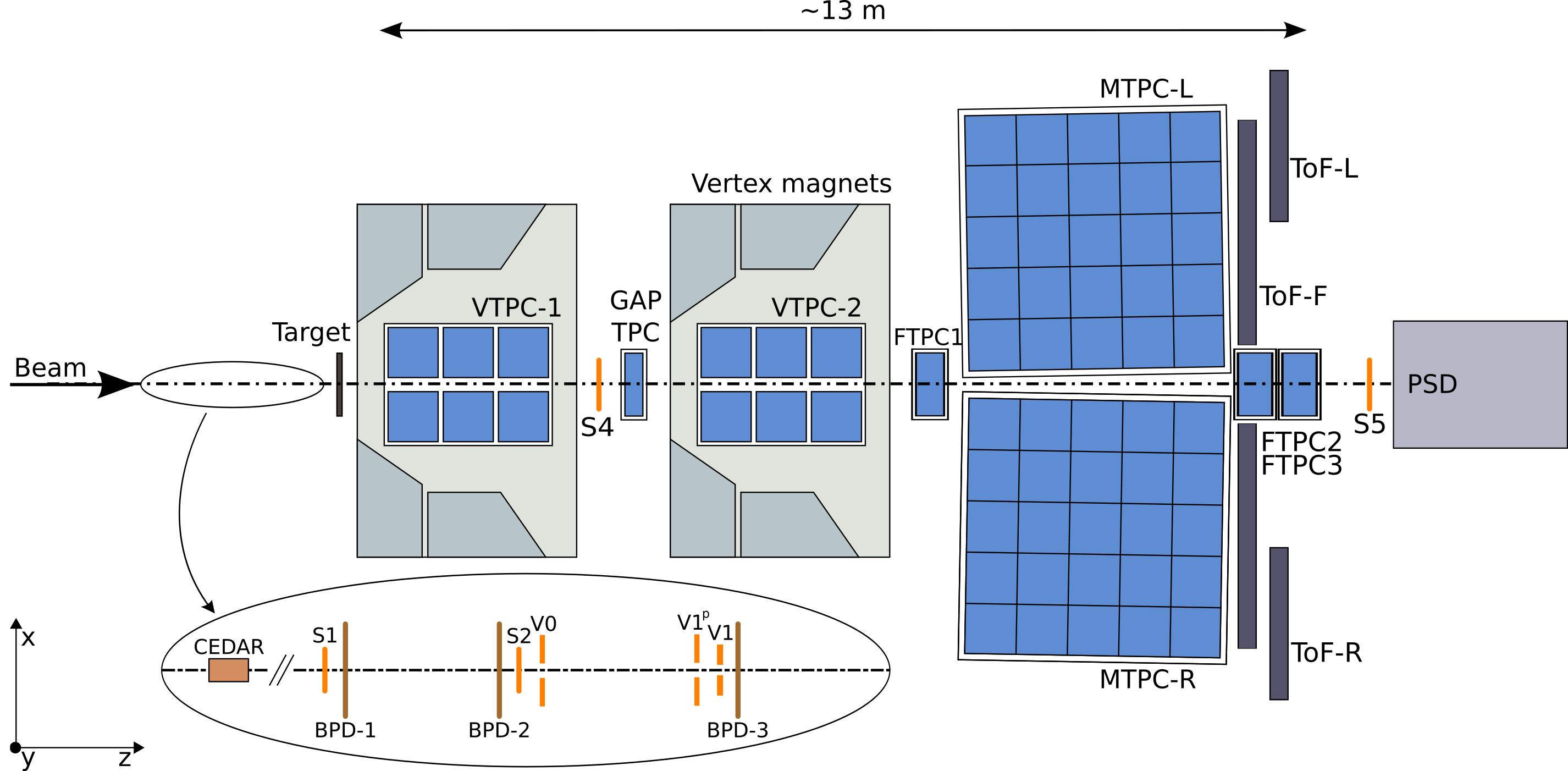}
 \caption{Schematic top-view layout of the NA61/SHINE experiment in the configuration used during the 2017 proton data taking. In 2016 the FTPCs were not present. The S5 scintillator was not used in this trigger configuration.
 }\label{fig:detectorConfiguration}
\end{figure*}

NA61/SHINE is a large-acceptance hadron spectrometer~\cite{na61detector}. Its Time Projection Chamber (TPC)-based tracking detectors are capable of reconstructing neutral hadrons that decay within detector acceptance, corresponding to a maximal decay length in the laboratory frame of approximately seven meters.

NA61/SHINE is located on the H2 beamline in Experimental Hall North 1 (EHN1) in CERN's North Area complex. The SPS provides the North Area with beams of primary 400 \gevc protons or ions with momenta in the range [13$A$ - 158$A$] \gevc. The protons can be directed into a production target to provide a beam of secondary hadrons in the range of 13 - 350 \gevc. These secondary beams contain a mixture of hadrons and leptons, and the desired beam particle species must be selected at the event level. Beam particle identification is performed by the Cherenkov Differential Counter with Achromatic Ring Focus (CEDAR)~\cite{cedar, cedar82}, located upstream of the NA61/SHINE spectrometer.

The components of the NA61/SHINE detector are shown in Fig.~\ref{fig:detectorConfiguration}. Eight TPCs act as the main tracking detectors and provide $dE/dx$ measurements for particle identification. The Vertex TPCs (VTPC-1 and VTPC-2) are located inside two superconducting vertex magnets, which provide up to 9 T$\cdot$m of maximum total bending power and enable track momentum measurement. A Time-of-Flight (ToF) system enables particle identification in selected regions of phase space. The Projectile Spectator Detector (PSD), a forward calorimeter, serves as a centrality detector. Three gaseous strip Beam Position Detectors (BPDs) measure incoming beam track trajectories. The BPDs are placed 29.5 m upstream (BPD1), 8.2 m upstream (BPD2), and 0.7 m upstream (BPD3) of the target. A straight line is fit to the three $(x,y)$ measurements made by the BPDs to represent the beam particle trajectory.

The beam trigger system, constructed from scintillators S1 \& S2, veto scintillators V0 \& V1 (scintillators with cylindrical holes centered on the beam), and the CEDAR detector, selects beam particles with acceptable trajectories and of the desired particle type. An interaction scintillator S4, placed downstream of the target, detects whether or not a significant angular scatter has occurred in the beam particle trajectory. 

Interactions of 120 \gevc protons and carbon were measured in 2016 and 2017 using a thin carbon target with dimensions 25\,mm (W) x 25\,mm (H) x 14.8\,mm (L) and density $\rho$ = 1.80 g/cm$^3$, corresponding to 3.1\% $\lambda$. Events were collected with the target removed to study neutral-hadron production outside of the target. The total number of recorded events and the number of events passing event preselection criteria (see Sec.~\ref{sec:V0Analysis}) are shown in Tab.~\ref{tab:selectedEventCounts}.

\begin{table*}[htbp]
\centering
\begin{tabular}{cccccccc}
Data Set  & Target-Inserted & Target-Inserted & Target-Removed & Target-Removed \\ 
& (Recorded) & (Selected) & (Recorded) & (Selected) \\
\hline
2016  & 3.6 M & 2.3 M & 0.20 M  & 0.08 M\\
2017  & 1.9 M & 1.6 M & 0.17 M  & 0.10 M
\end{tabular}
\caption[Event Counts]{The number of recorded and selected target-inserted (target-removed) events for the 2016 and 2017 data samples.}
\label{tab:selectedEventCounts}
\end{table*}

The detector configuration was significantly altered in 2017, resulting in different acceptances and measurement capabilities for the two data sets. For the neutral-hadron analyses, the most significant difference is the magnetic field strength, which was reduced by half in 2017 due to the addition of new forward-tracking TPCs~\cite{Rumberger:FTPCPaper}. The addition of these TPCs has a significant impact on the charged hadron analysis, which will be presented in a separate publication.

\section{Data Reconstruction \& Simulation}\label{sec:Reconstruction}

The recorded data sets were reconstructed using the NA61/SHINE "Legacy" reconstruction chain, which includes a \vo finder and Minuit-based \vo fitter~\cite{James:1994vla}. The performance of this reconstruction chain has been described in previous publications~\cite{na61detector}. A \GeantFour-based~\cite{Agostinelli:2002hh,Allison:2006ve,Allison:2016lfl} detector description is used to simulate the passage of particles through the NA61/SHINE detectors and evaluate reconstruction efficiency and detector acceptance. The particular physics list used to calculate acceptance and reconstruction corrections was the FTFP\_BERT physics list (\GeantFourVersion). FTFP\_BERT uses the Bertini Cascade model~\cite{Wright:2015xia} for hadronic interactions below 5 \gev and the Fritiof model~\cite{Uzhinsky:2013hea} for interactions above 5 \gev. The Monte Carlo correction calculation procedure will be described in Section~\ref{sec:V0Analysis}.
 
\section{Neutral-Hadron Multiplicity Analysis} \label{sec:V0Analysis}

The neutral-hadron analysis includes reconstructing neutral decay vertices, referred to as \vos, applying selection criteria to these \vos, fitting invariant mass spectra, and calculating identified multiplicities. Double-differential multiplicity results are reported as a function of neutral-hadron production angle $\theta$ and momentum magnitude $p$. Neutral weakly-decaying hadrons included in the \vo analysis are \kos (with $\kos \to \pip \; \pim$), \lam (with $\lam \to \p \; \pim$) and \alam (with $\alam \to \ap \; \pip$). Candidate \vos are identified by pairing all possible combinations of positively-charged and negatively-charged TPC tracks in an event and selecting pairs with a distance of closest approach of less than 5\,cm. Invariant mass, center-of-mass kinematics, and laboratory kinematics are calculated for such compatible track pairs.

Event and track selection for the neutral-hadron analyses follow a similar methodology to previous NA61/SHINE measurements for \piCnoE and \piBe~\cite{Aduszkiewicz:2019hhe}. Selection criteria used in this analysis are discussed in the following subsections.

\subsection{Event Preselection}

Three selection criteria are applied at the event level prior to track selection.

\begin{itemize}
\item{Beam Divergence Cut (BPD Cut)}
\end{itemize}

To mitigate systematic effects related to large beam divergence, a cut is applied to each measured beam particle trajectory. Beam tracks with significant angle will miss the S4 scintillator and cause an interaction trigger, even if no significant interaction occurred. The BPD cut ensures that the unscattered trajectory of each beam track points to the S4 scintillator.

\begin{itemize}
\item{Well-Measured Beam Trajectory Cut (BPD Status Cut)}
\end{itemize}

The BPD status is an indicator of how well an incoming beam particle trajectory is measured. Any one of the three BPDs may report an error during the clusterization and fitting process due to transient noise in the detector or another ionizing particle passing through the detector simultaneously. The BPD status cut ensures that either all three detectors measured the six coordinates of a particle's trajectory and a straight line fit converged, or that two of the detectors reported satisfactory measurements and a straight line fit converged. BPD3 is required to have a single well-measured cluster, ensuring that no significant scatter occurred upstream of BPD3.

\begin{itemize}
\item{Off-Time Beam Particle Cut (WFA Cut)}
\end{itemize}

The final event-level cut reduces systematic effects associated with beam intensity. The Waveform Analyzer (WFA) records signals in the trigger scintillators near the triggered event, including those from beam particles not associated with the interaction trigger. These are known as off-time beam particles. The arrival of a subsequent beam particle closely-spaced in time may hit the S4 scintillator and appear to be a non-interaction. In addition, off-time beam particles may interact in the target. If the off-time particle arrives several hundred nanoseconds after the triggering particle, off-time tracks may be reconstructed to the event main vertex. In order to reduce these effects, a WFA cut of 0.8 $\mu$s was used.

\begin{figure*}[t]
  \centering
  \includegraphics[width=0.35\textwidth]{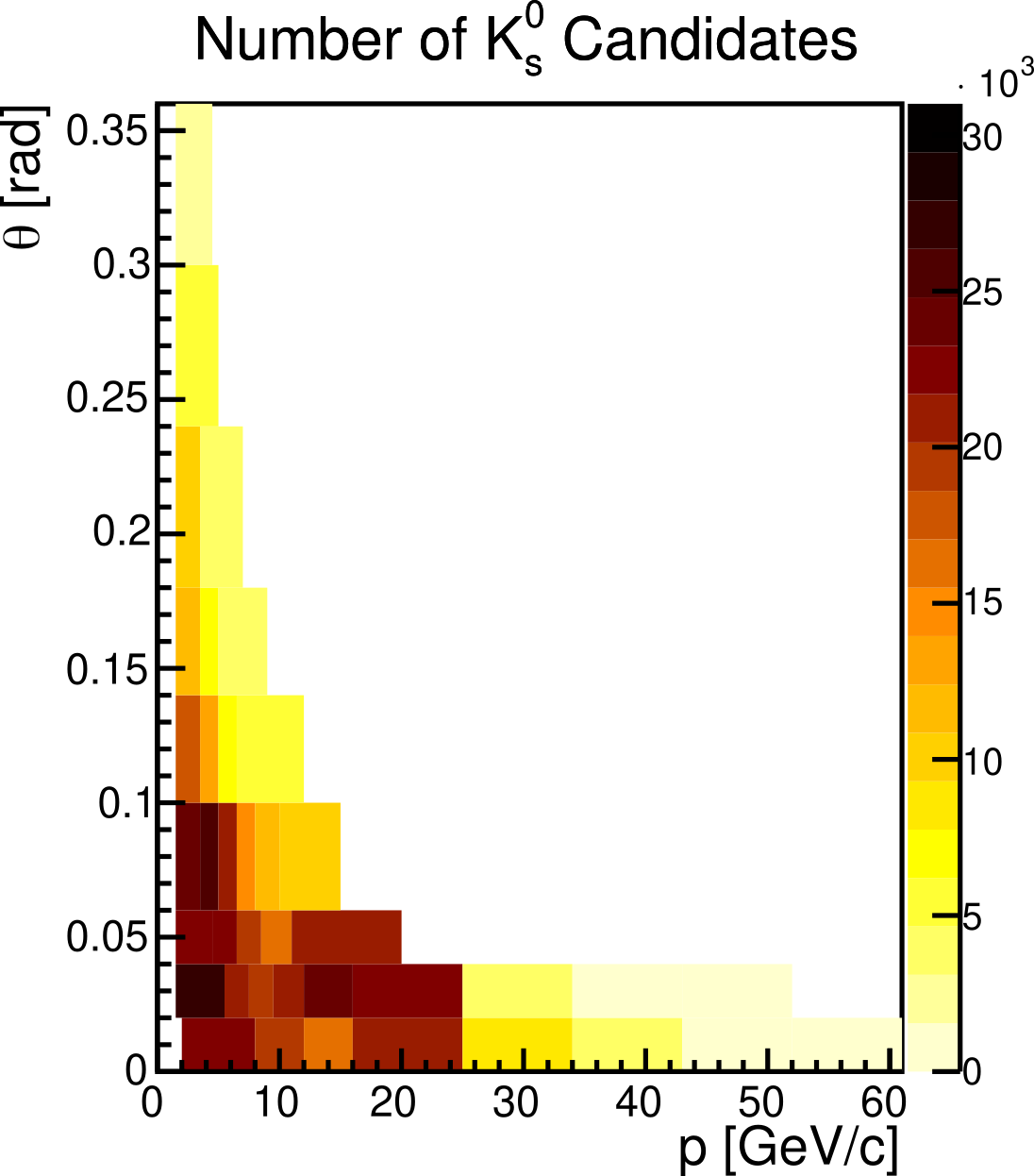}
  \includegraphics[width=0.35\textwidth]{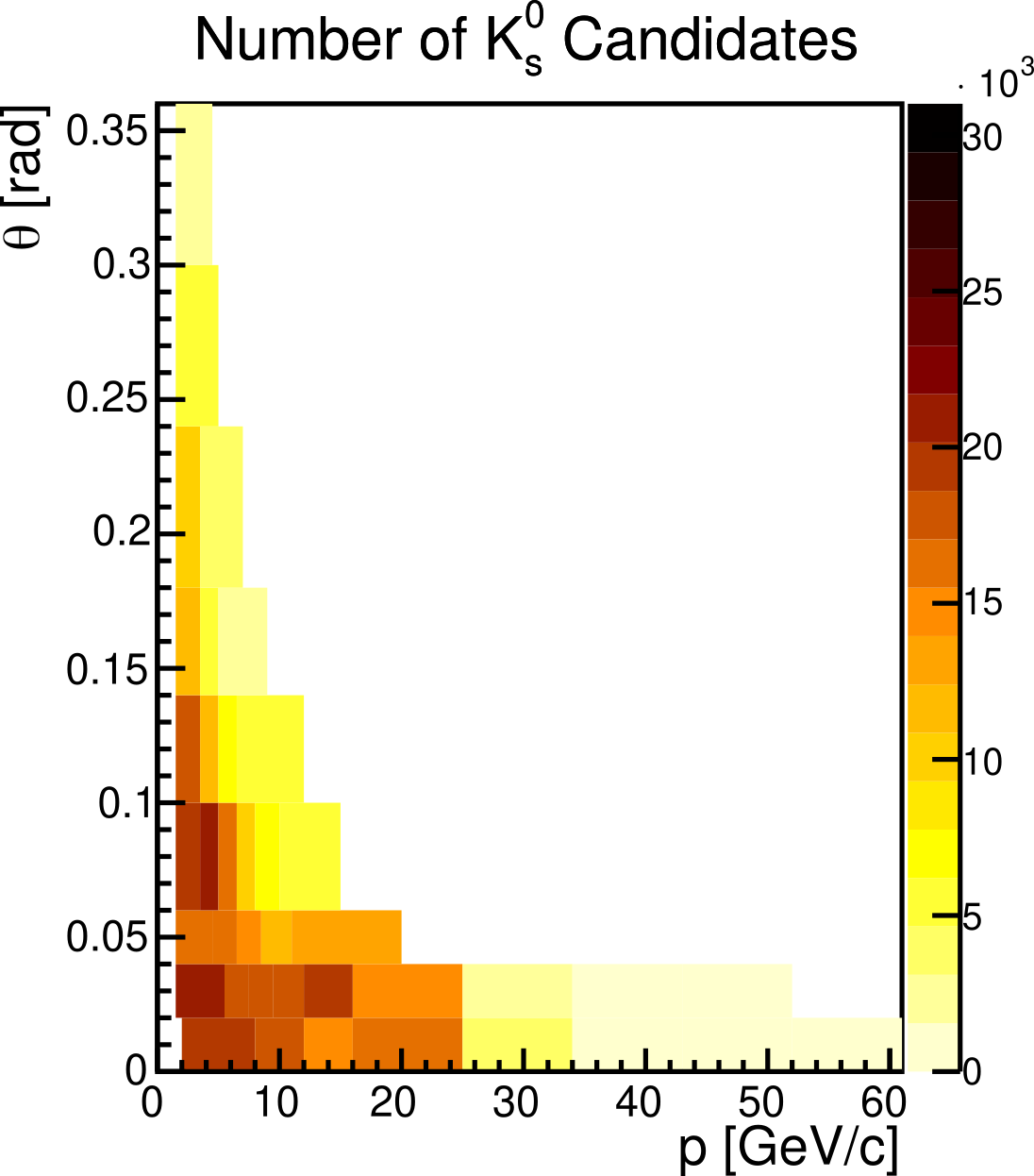}
\caption{Selected \vo candidates for \kos analysis. Left: 2016 analysis kinematic bin occupancy. Right: 2017 analysis kinematic bin occupancy. }\label{fig:occupancyK0S}
\end{figure*}

For spectra analysis, only interaction trigger events are considered.
After the described selection cuts, 2.2 M (2016) and 1.6 M (2017) target-inserted and 0.08 M (2016) and 0.1 M (2017) target-removed events were selected (see Tab.~\ref{tab:selectedEventCounts}). Differences in the target-inserted and target-removed ratios between the two years are simply due to different amounts of beam time being devoted to target-removed event collection.

\subsection{Selection of \vo Candidate Tracks}

 Cuts intended to improve sample purity, which are specific to the particular neutral-hadron species being analyzed, are discussed below. These cuts remove \vos that likely did not originate from the neutral particle of interest.

\begin{figure*}[t]
  \centering
  \includegraphics[width=0.35\textwidth]{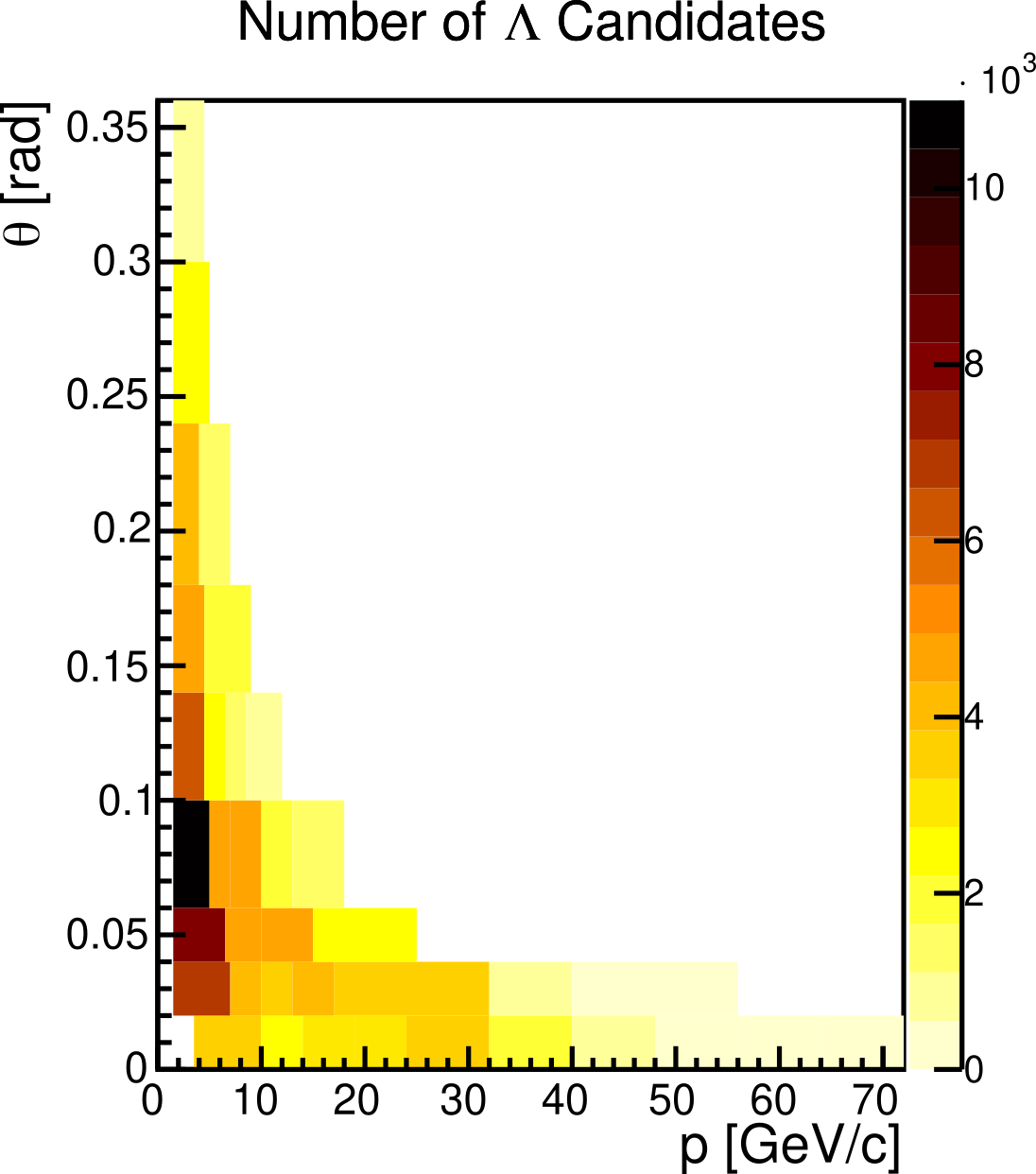}
  \includegraphics[width=0.35\textwidth]{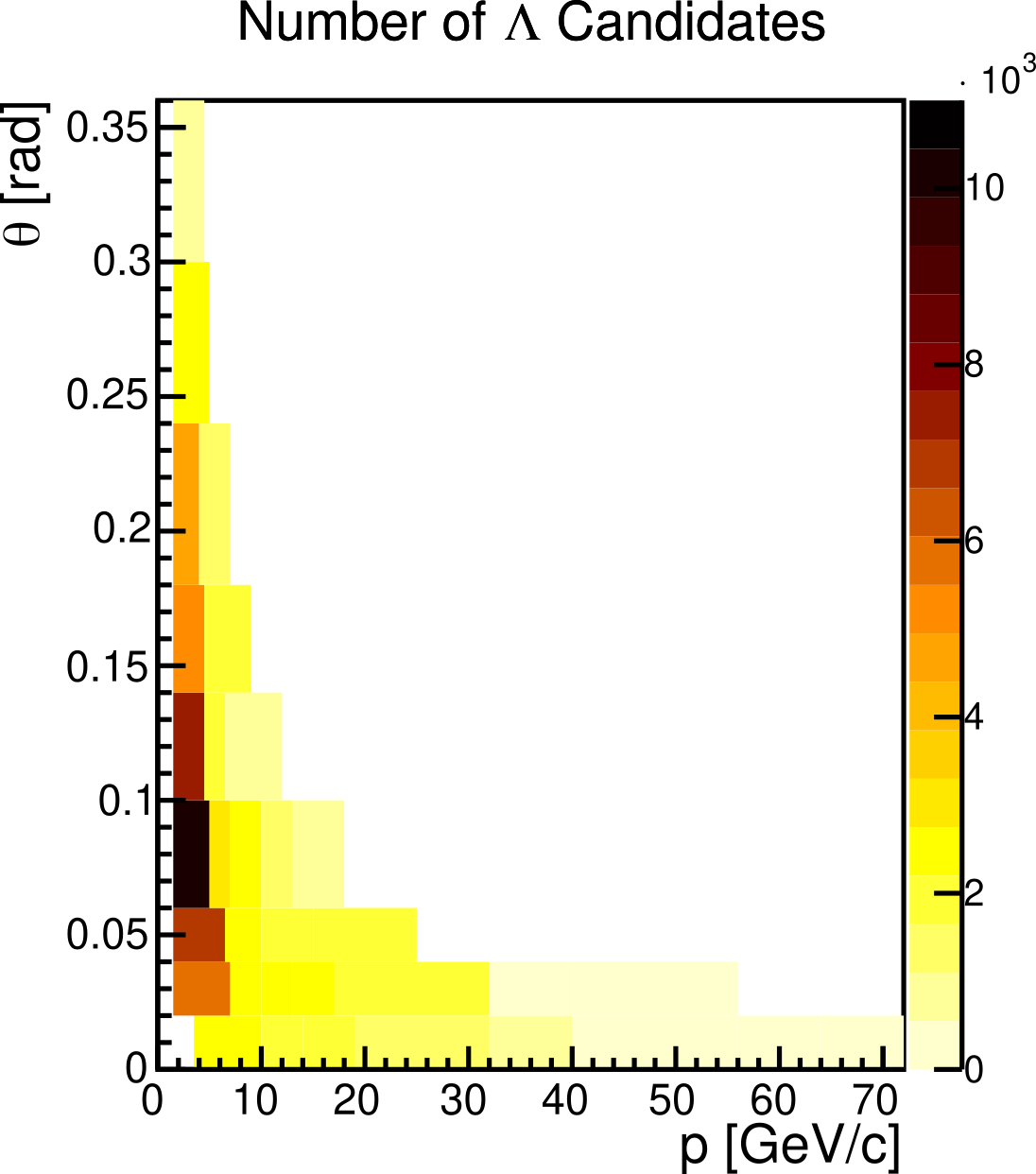}
\caption{Selected \vo candidates for \lam analysis. Left: 2016 analysis kinematic bin occupancy. Right: 2017 analysis kinematic bin occupancy. }\label{fig:occupancyLam}
\end{figure*}

\begin{itemize}
\item{\vo Topological Cuts}
\end{itemize}

Several cuts are common to the analysis of each neutral-hadron species. Selected \vos must be separated from the primary vertex by at least 3.5 cm, in order to remove fake \vo contributions from the primary interaction. Charged child tracks must have at least 12 total point measurements in the VTPCs (VTPC1 + VTPC2), in order to reliably reconstruct track momenta. The \vo impact parameter, the distance between the extrapolated neutral track position and the event primary vertex at the target plane, must be less than 4 cm in the bending plane and 2 cm in the non-bending plane. Finally, to reject converted photons, a cut on the transverse momentum of the decay in a co-moving frame with the \vo is imposed: $p_T^+ + p_T^- > 30$ \mevc.

\begin{figure*}[t]
  \centering
  \includegraphics[width=0.35\textwidth]{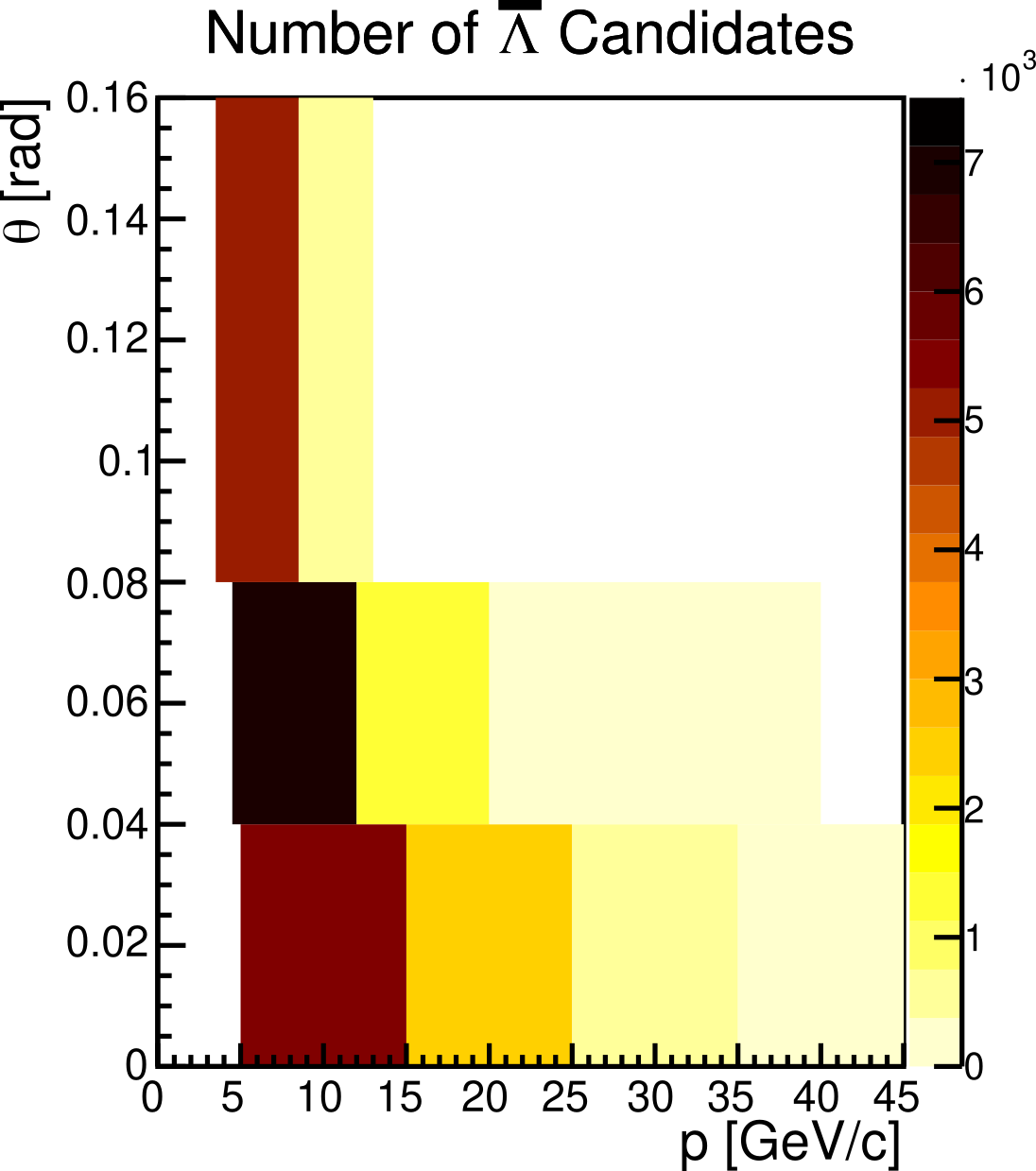}
  \includegraphics[width=0.35\textwidth]{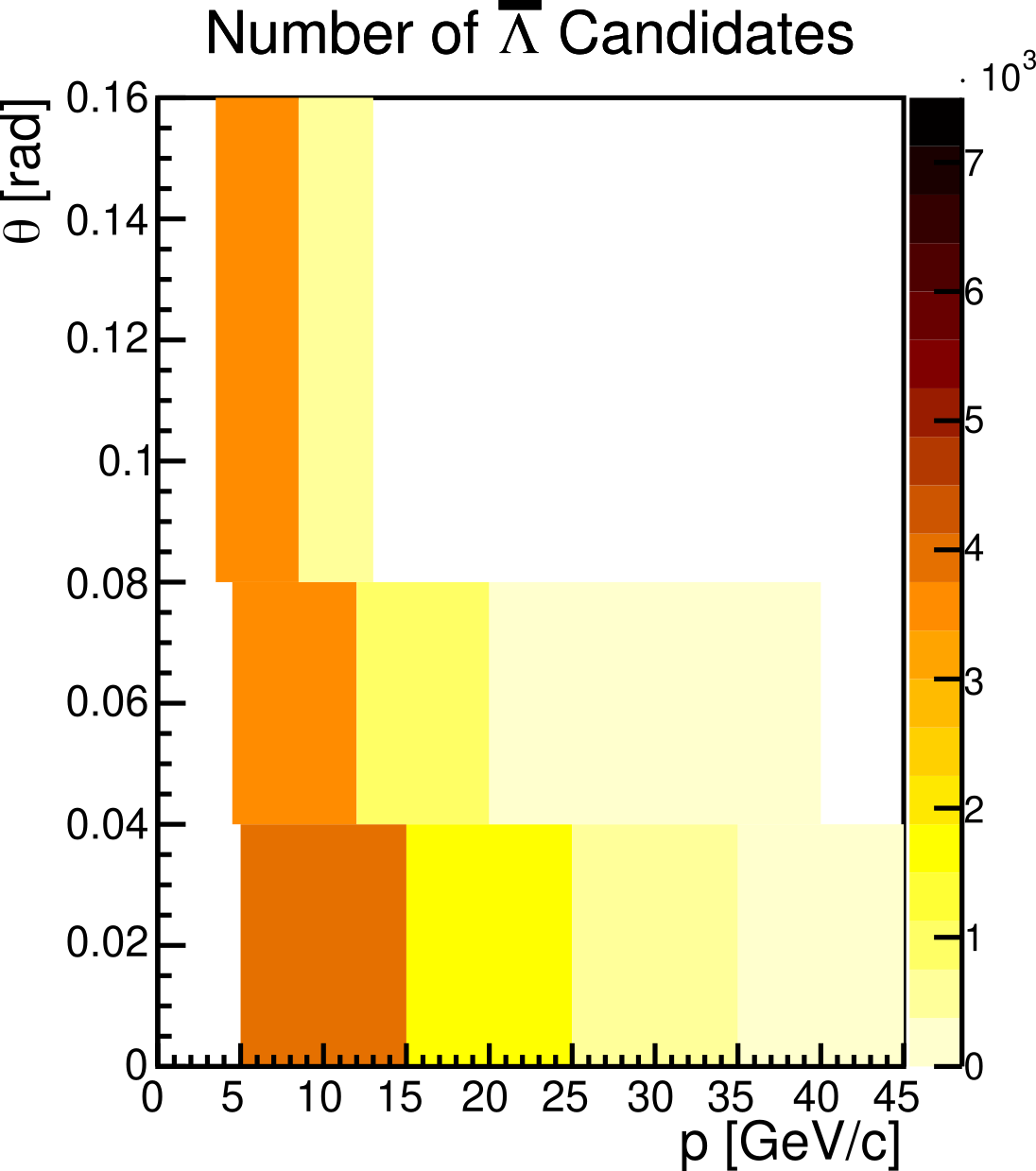}
\caption{Selected \vo candidates for \alam analysis. \emph{Left}: 2016 analysis kinematic bin occupancy. \emph{Right}: 2017 analysis kinematic bin occupancy. }\label{fig:occupancyALam}
\end{figure*}

\begin{itemize}
\item{Purity Cuts}
\end{itemize}

The remaining cuts are designed to increase the purity for the particular hadron of interest and are therefore specific to each species. A decay hypothesis is assumed: For \kos, the positively-charged and negatively-charged tracks are assumed to be \pipm, while for \lam (\alam) the positively-charged track is assumed to be a $p$ (\pip) and the negatively-charged track is assumed to be a \pim ($\bar{p}$). Protons, antiprotons, and charged pions are assigned masses corresponding to their current best-fit values~\cite{PDG}. For Lorentz factor calculation, \kos tracks are assigned masses of 497.6 \mev and \lam \& \alam are assigned masses of 1115.6 \mev~\cite{PDG}. 

Restrictions are imposed on the angle formed by the child tracks in the decay frame and the \vo direction of travel. For \kos, the allowed angular regions are $\cos{\theta^{+*}} \in [-0.9,0.7]$ and $\cos{\theta^{-*}} \in [-0.7,0.9]$, while for \lam (\alam) the allowed angular regions are $\cos{\theta^{+*}} \in [-0.7,0.9]$ and $\cos{\theta^{-*}} \in [-0.9,0.7]$ ($\cos{\theta^{+*}} \in [-0.9,0.7]$ and $\cos{\theta^{-*}} \in [-0.7,0.9]$).

The range of invariant mass $m_\textrm{inv} = \sqrt{m^2_+ + m^2_- + 2 (E_+ E_- - \Vec{p_+} \cdot \Vec{p_-})}$ is restricted for each analysis. For \kos the allowed range is $m_\textrm{inv} \in [0.4,0.65]$ \gev, while for \lam \& \alam  the allowed range is $m_\textrm{inv} \in [1.09,1.215]$ \gev. This mass range allows for reasonable background shape fitting around the signal region of interest.

\begin{figure*}[t]
  \centering
  \includegraphics[width=0.45\textwidth]{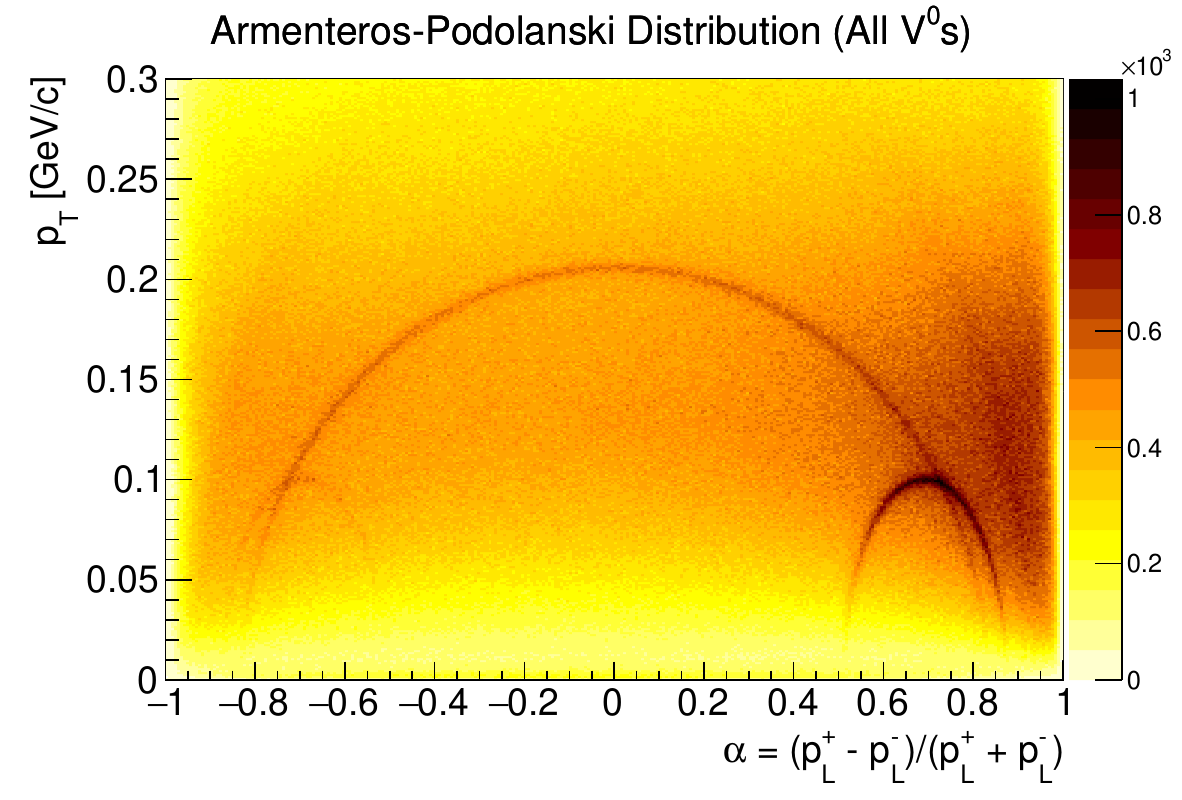}
  \includegraphics[width=0.45\textwidth]{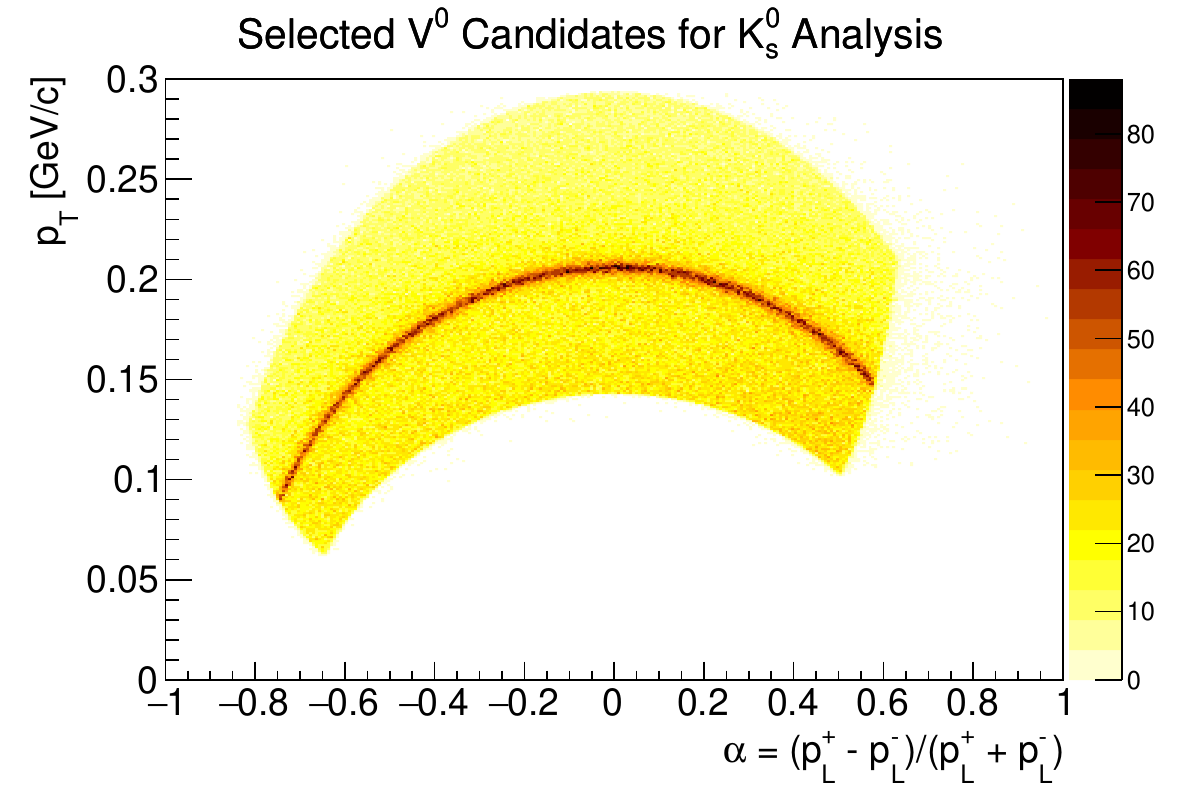}
  \includegraphics[width=0.45\textwidth]{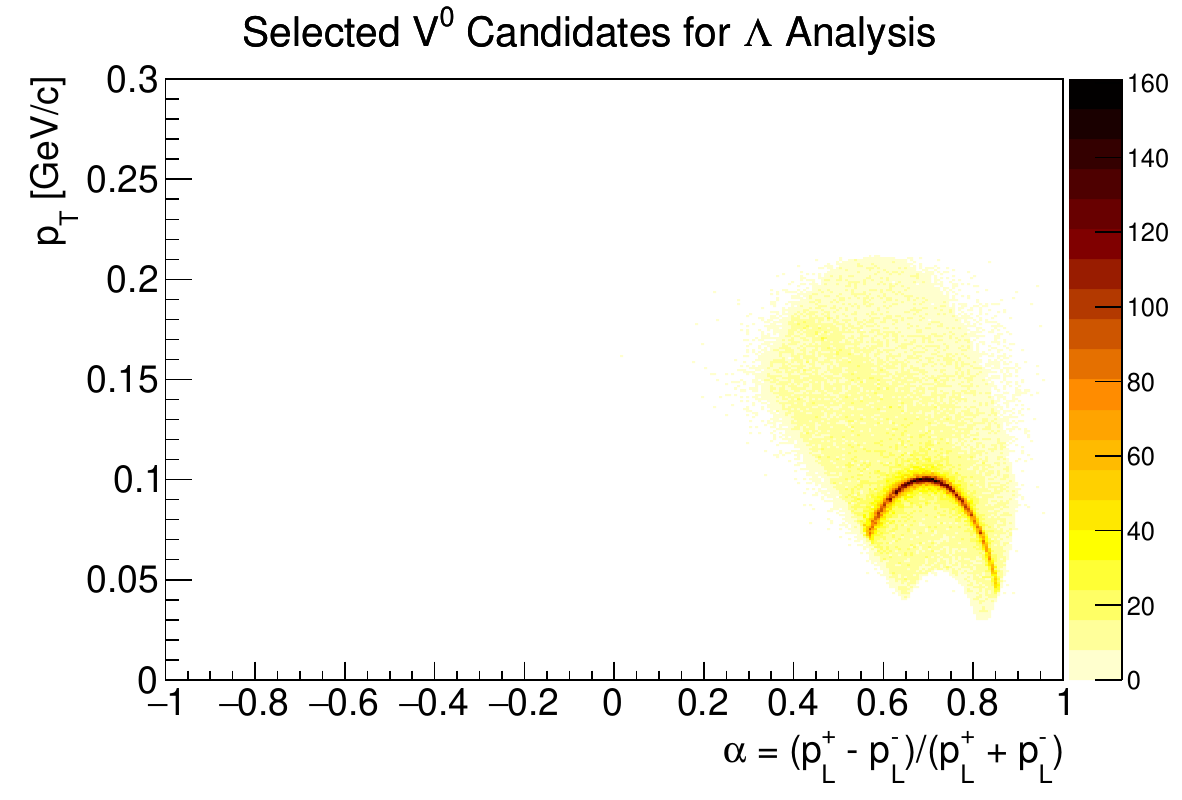}
  \includegraphics[width=0.45\textwidth]{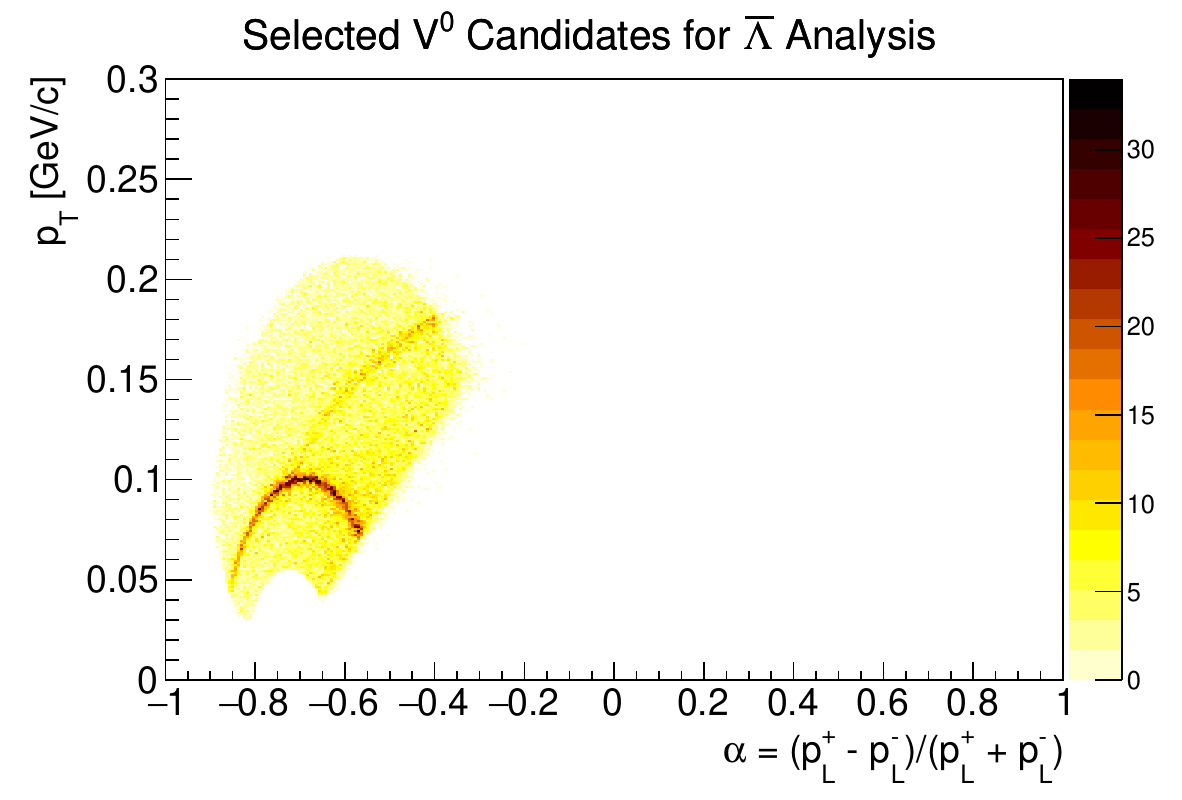}
\caption{Armenteros--Podolanski distributions from the 2016 data set before and after applying selection criteria for the \kos, \lam and \alam analyses.}\label{fig:armenterosPodolanski}
\end{figure*}

A cut on decay product $dE/dx$ is imposed, significantly reducing \pim (\pip) contamination in the \lam (\alam) analysis. This cut is also applied to the \kos analysis. In each analysis, the measured $dE/dx$ of each decay product is examined. If one of the decay products $dE/dx$ differs from the expected decay product $dE/dx$ by more than 15\%, the \vo is omitted from the analysis.

Finally, a cut on normalized proper lifetime is imposed. The \kos lifetime is assigned as $c\tau = 2.68$ cm, while for \lam \& \alam $c\tau = 7.89$ cm~\cite{PDG}. The reconstructed lifetime cut is $c\tau > 0.25 c \; \tau_\textrm{PDG}$.

\begin{table*}[htbp]
\centering
\begin{tabular}{cccccccc}
Data Set  & Selected \kos Candidates & Selected \lam Candidates & Selected \alam Candidates \\
\hline
2016  & 536 K (22 K) & 120 K (3.6 K) & 45 K (1.3 K) \\
2017  & 430 K (17 K) & 90 K (3.1 K) & 35 K (1.2 K)
\end{tabular}
\caption[Candidate \vos Passing Selection Cuts]{The number of candidate target-inserted (target-removed) \vos passing selection cuts for the 2016 and 2017 data samples.}
\label{tab:selectedV0Counts}
\end{table*}

\subsection{Armenteros--Podolanski Distributions}

The impacts of the selection cuts can be examined using Armenteros--Podolanski distributions, which plot transverse momentum $p_T$ as a function of longitudinal momentum asymmetry $\alpha$ in a co-moving frame with the \vo~\cite{armenterosPodolanski}: 

\begin{equation}
\alpha = \frac{p^+_L - p^-_L}{p^+_L + p^-_L}.
\end{equation}

Here $p^{\pm}_L = p^{\pm}\cos{\theta^{\pm}}$, $p_T = p^{+}_T + p^{-}_T$, and $p^{\pm}_T = p^{\pm}\sin{\theta^{\pm}}$. $\vec{p^\pm}$ are the positively-charged and negatively-charged child track three-momenta in the neutral hadron's rest frame. Figure \ref{fig:armenterosPodolanski} shows the Armenteros--Podolanski distributions before and after applying track selection cuts to the 2016 data set.  

\subsection{Invariant Mass Distribution Fits}

After applying cuts, remaining candidate \vos are collected and sorted into kinematic analysis bins. The number of \vo candidates in each kinematic bin after applying all selection cuts can be seen in Figs.~\ref{fig:occupancyK0S}--~\ref{fig:occupancyALam}. In each kinematic bin, an invariant mass spectrum fit is performed in order to extract the number of signal \vos. 

\begin{figure*}[t]
  \centering
  \includegraphics[width=0.32\textwidth]{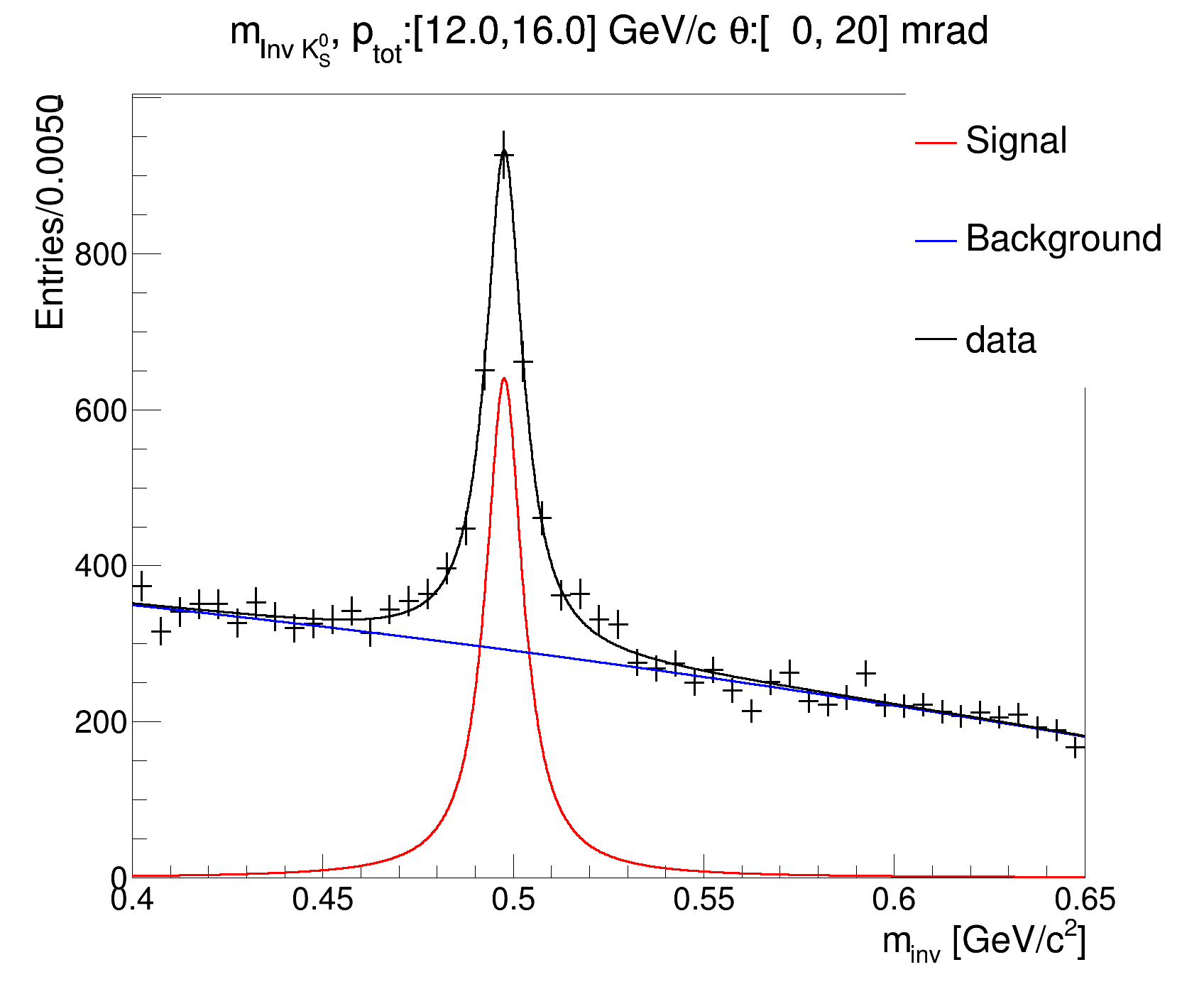}
  \includegraphics[width=0.32\textwidth]{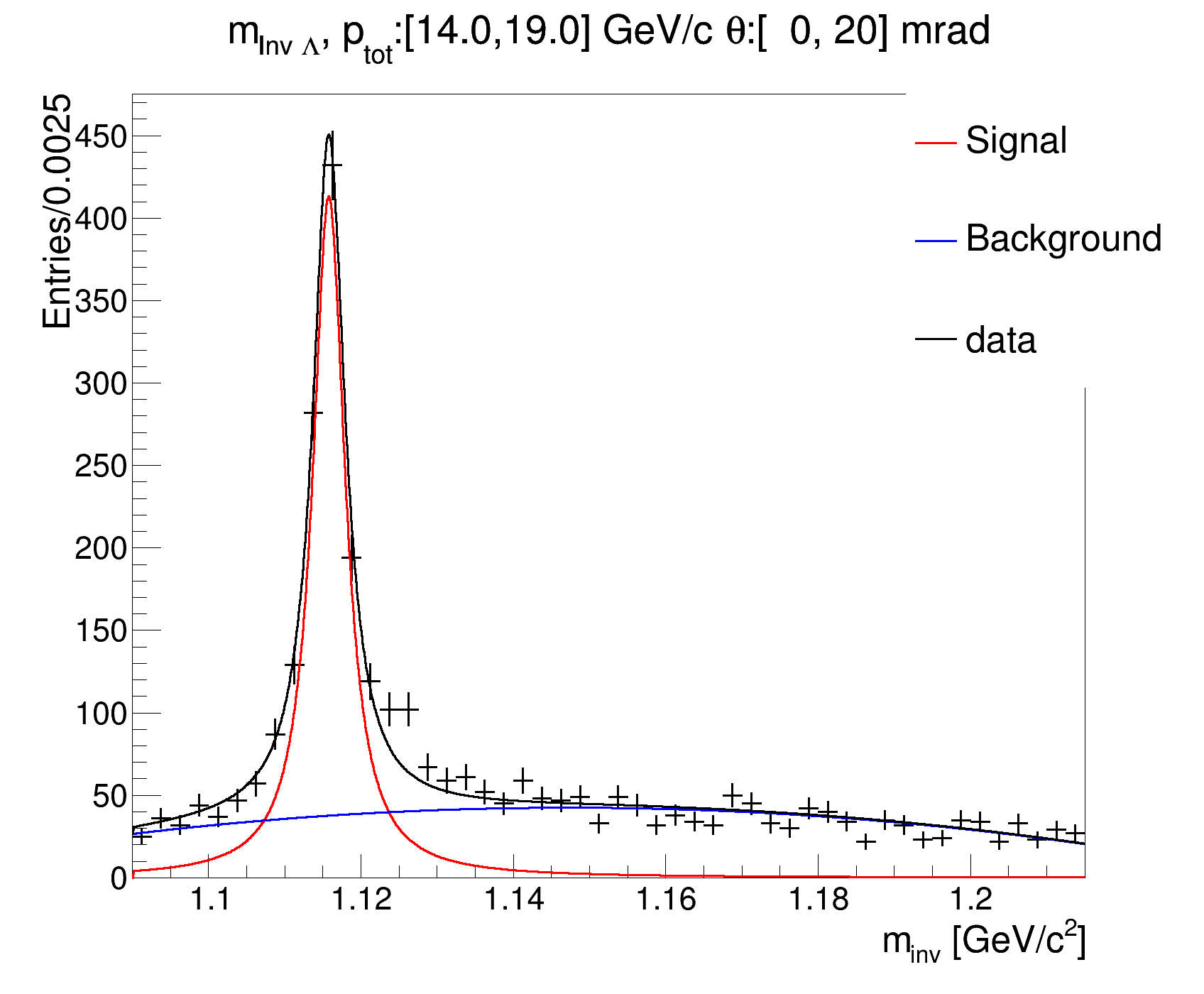}
  \includegraphics[width=0.32\textwidth]{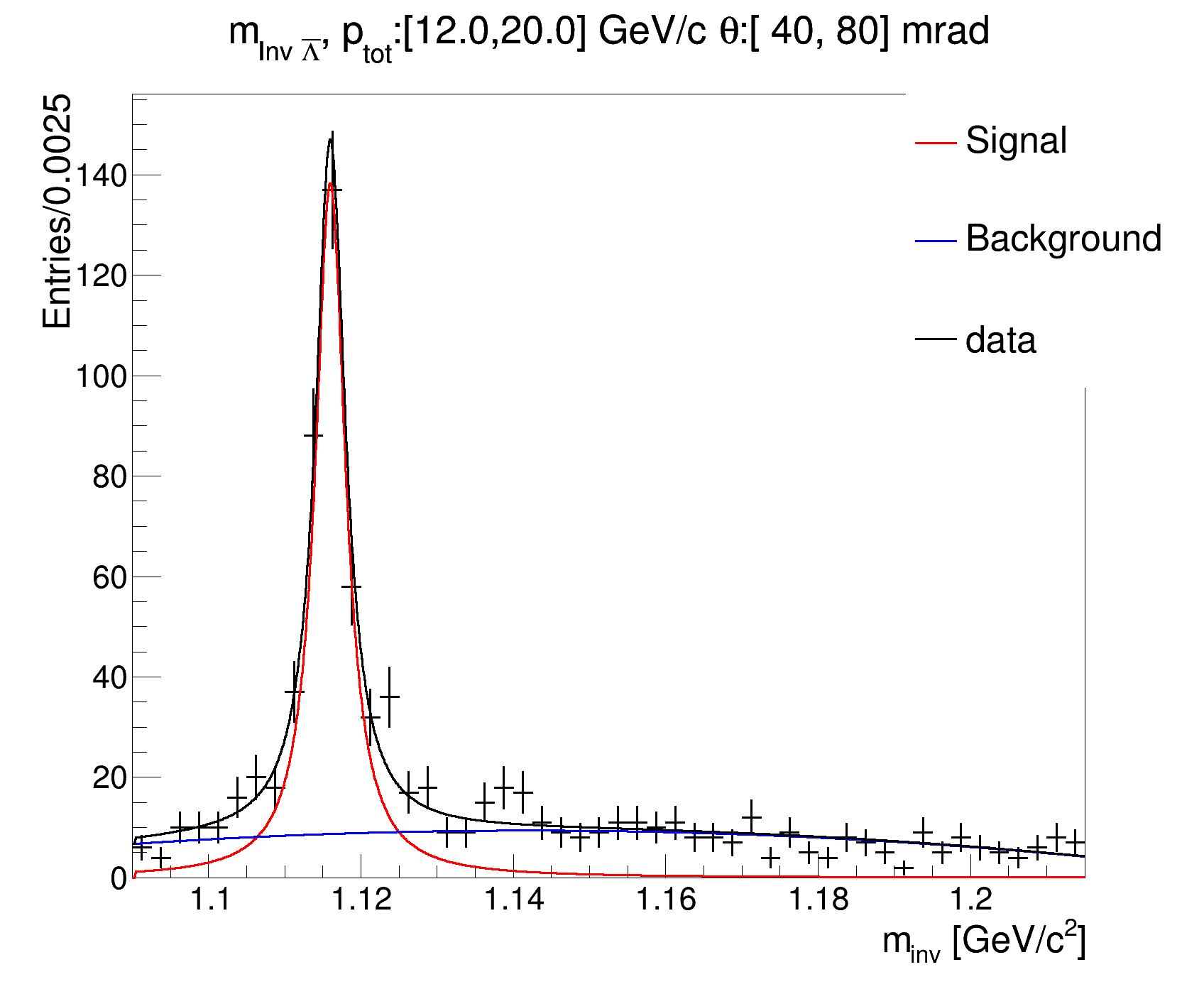}
\caption{Example invariant mass distribution fits for the \kos (left), \lam (center), and \alam (right) analyses.}\label{fig:invariantMassFits}
\end{figure*}

A representative invariant mass distribution fit for one kinematic bin from each analysis can be seen in Fig.~\ref{fig:invariantMassFits}.

The fits were performed by minimizing a continuous likelihood function,

\begin{equation}
\log L = \sum_{\textrm{\vo candidates}} \log{F(m_\textrm{inv};\vec{\theta})},
\end{equation}

where $\vec{\theta}$ are the fit parameters and

\begin{equation}
F(m_\textrm{inv},\vec{\theta}) = c_\textrm{s} f_\textrm{s} (m_\textrm{inv};\vec{\theta}_\textrm{s}) + (1-c_\textrm{s}) f_\textrm{bg} (m_\textrm{inv};\vec{\theta}_\textrm{bg}).
\end{equation}

Here $c_s$ is the fraction of \vos considered to be signal \vos. The background function $f_\textrm{bg} (m_\textrm{inv};\vec{\theta}_\textrm{bg})$ is a third-order polynomial.

The signal model used is a Lorentzian:

\begin{equation}
f_\textrm{s}(m;m_0,\Gamma) = \frac{1}{\pi \Gamma} \frac{\Gamma^2}{(m-m_0)^2 + \Gamma^2}.
\end{equation}

Here $m$ and $m_0$ are the invariant mass and offset from the accepted best-fit value, respectively, and $\Gamma$ describes the distribution width. Central invariant masses were allowed to deviate from the known particle masses, to allow for momentum mis-reconstruction in certain regions of phase space. 

The signal yield for each kinematic bin is calculated using the signal fraction $c_s$ and the total number of \vos in the bin:

\begin{equation}
y^{\textrm{raw}}_i= (c_s N_\textrm{\vo candidates})_i,
\end{equation}

where $i$ corresponds to the kinematic bin number.

\subsection{Correction Factors}

A bin-dependent Monte Carlo correction factor was calculated in order to estimate the number of true signal \vos from the raw measured yields. This factor corrects for detector acceptance, reconstruction efficiency, S4 efficiency, selection efficiency, feed-down corrections (a correction for neutral hadrons produced via weak decay), and the measured decay channel's branching ratio. The correction factor is defined as

\begin{equation}
c_i = \frac{N(\textrm{simulated signal \vos)}}{N(\textrm{reconstructed fit signal \vos)}} =
  c_{\textrm{acc.}} \times c_{\textrm{sel.}} \times c_{\textrm{rec. eff.}} \times c_{\textrm{fd}} \times c_{\textrm{br.}},
\end{equation}

where $i$ indicates the kinematic bin. $c_{\textrm{acc.}}$ is the correction associated with acceptance cuts, $c_{\textrm{sel.}}$ is the correction associated with track quality cuts, $c_{\textrm{rec. eff.}}$ is the correction associated with reconstruction efficiency, $c_{\textrm{fd}}$ is the estimated feed-down correction for \lam and \alam originating from weak decays of $\Xi$ and $\Omega$ baryons, and $c_{\textrm{br.}}$ is the branching fraction correction, as we only measure one decay channel for each neutral hadron species. The correction factors were obtained using the FTFP\_BERT physics list. 

\subsection{Neutral-Hadron Multiplicity Measurements}\label{sec:V0MultiplicityMeasurements}

The differential production multiplicity is defined as the average number of particles produced in a given kinematic bin $i$ per unit momentum per unit angle in a production interaction. This can be expressed using the trigger and production cross sections $\sigma_\textrm{trig}$ and $\sigma_\textrm{prod}$, which correspond to the probability of a beam particle causing a trigger and the probability of a beam particle causing a production interaction, respectively:

\begin{equation}\label{eq:multiplicity}
\frac{d^2n_i}{dpd\theta} = \frac{c_i \; \sigma_\textrm{trig}}{\Delta p\Delta\theta \; \sigma_\textrm{prod}} \frac{y_i}{N_\textrm{trig}}.
\end{equation}

Here $\Delta p \Delta \theta$ is the size of kinematic bin $i$, $y_i$ is the raw fit yield in the kinematic bin, and $N_\textrm{trig}$ is the total number of accepted events. After applying selection cuts to the target-removed data samples, invariant mass fits in each bin reported a negligible amount of \kos, \lam, and \alam. Target-removed subtraction was therefore not performed.

\begin{figure*}[t]
  \centering
  \includegraphics[width=\textwidth]{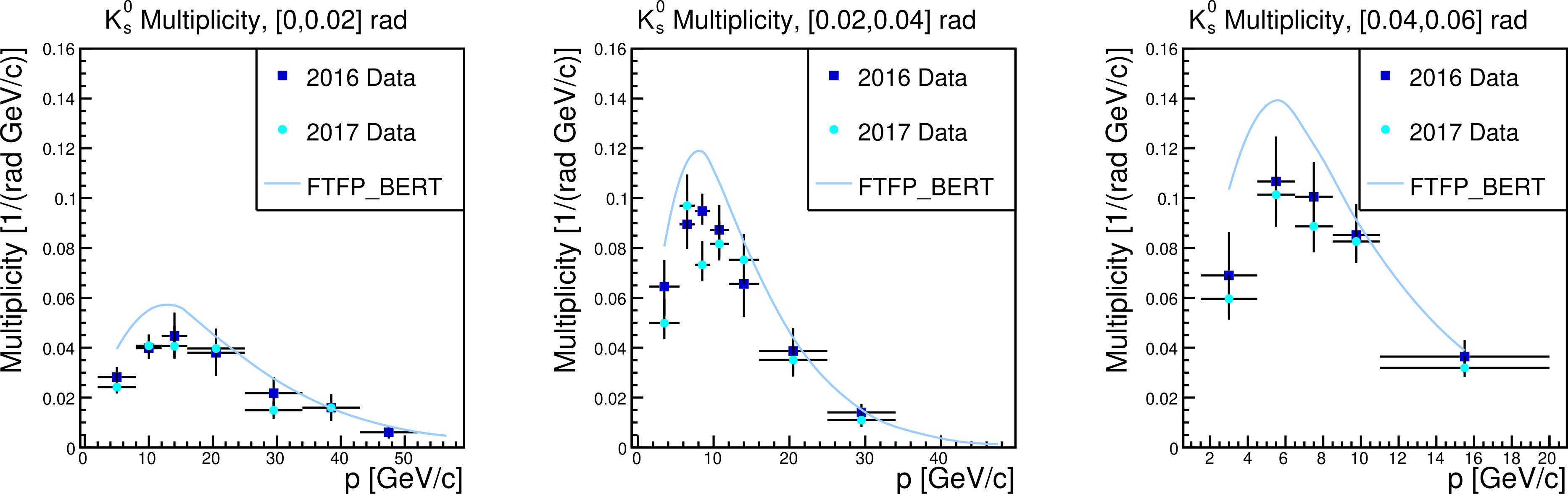}
\caption{\kos multiplicity measurements from 2016 and 2017 data sets for three angular bins.  Uncertainties reflect total uncorrelated uncertainty (statistical and uncorrelated systematic) for the independent analyses.}\label{fig:neutralComparisonK0S}
\end{figure*}

\begin{figure*}[t]
  \centering
  \includegraphics[width=\textwidth]{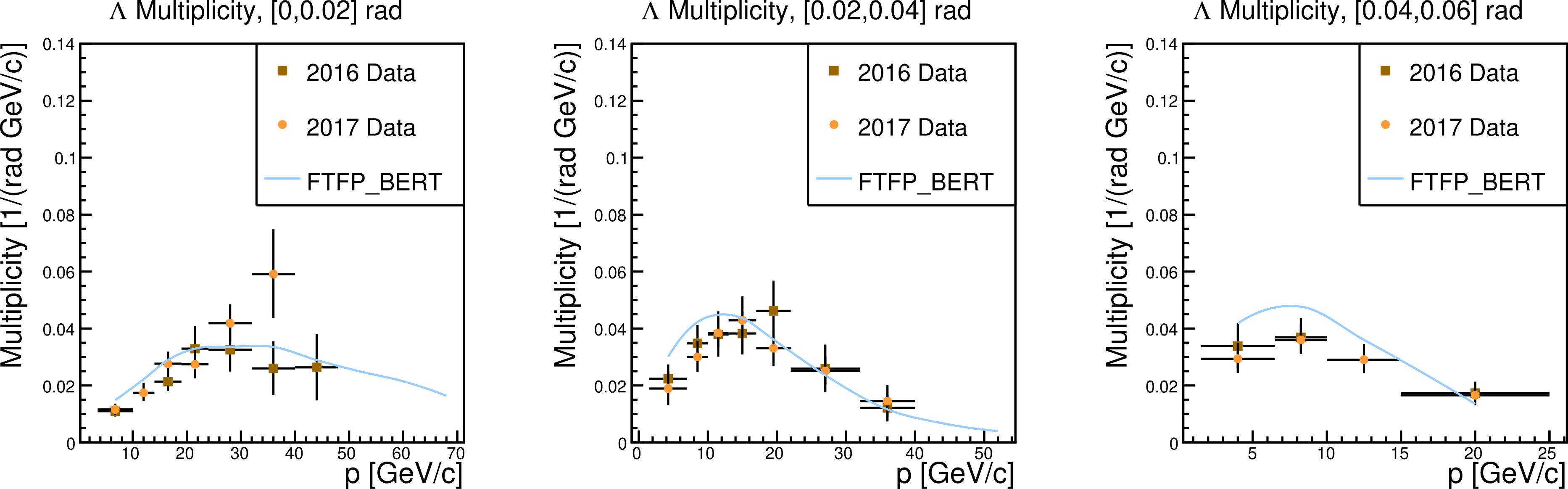}
\caption{\lam multiplicity measurements from 2016 and 2017 data sets for three angular bins. Uncertainties reflect total uncorrelated uncertainty (statistical and uncorrelated systematic) for the independent analyses.}\label{fig:neutralComparisonLam}
\end{figure*}

\begin{figure*}[t]
  \centering
  \includegraphics[width=\textwidth]{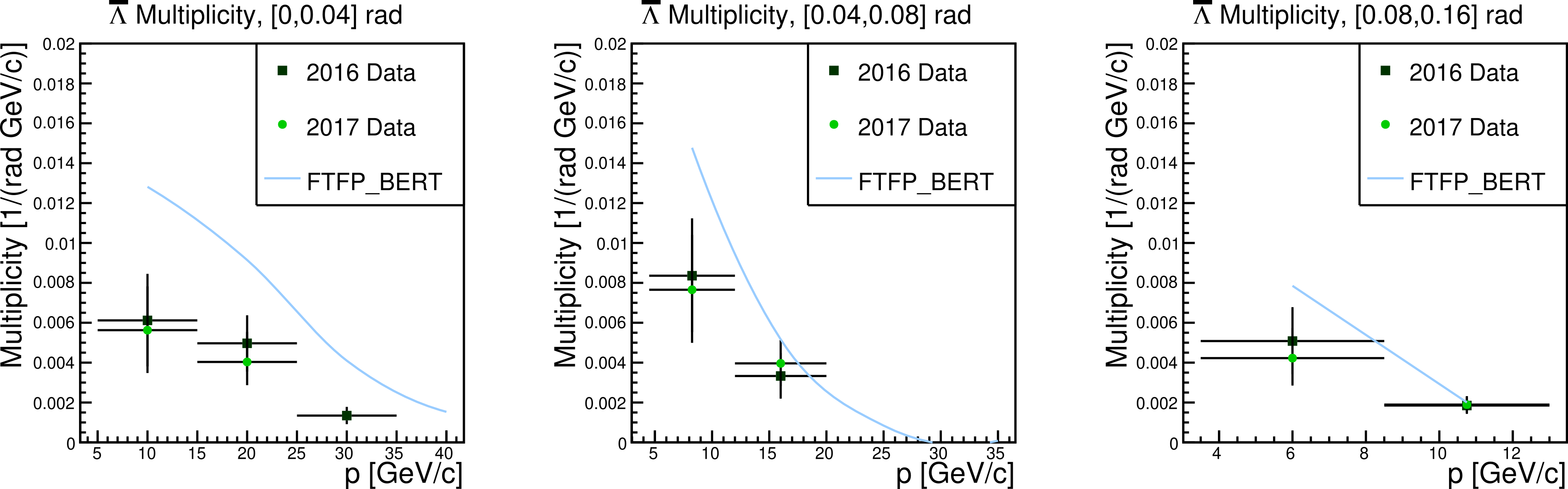}
\caption{\alam multiplicity measurements from 2016 and 2017 data sets for three angular bins.  Uncertainties reflect total uncorrelated uncertainty (statistical and uncorrelated systematic) for the independent analyses.}\label{fig:neutralComparisonALam}
\end{figure*}

Production multiplicities in selected regions of phase space for \kos, \lam, and \alam are presented in Figs.~\ref{fig:neutralComparisonK0S}--~\ref{fig:neutralComparisonALam}. Comparisons of the independent 2016 and 2017 multiplicity measurements show agreement within 1$\sigma$ for the majority of the measurements. A combined measurement, taking into account correlated and uncorrelated systematic uncertainties, will be presented in Section~\ref{sec:combinedMultiplicities}.

\section{Systematic Uncertainties of 2016 and 2017 Analyses}

Systematic uncertainties from several effects were considered and their effects were evaluated independently for the 2016 and 2017 analyses. This section will detail sources of uncertainty considered and show the individual contributions to total systematic uncertainty. 

\begin{figure*}[!ht]
  \centering
  \includegraphics[width=0.3\textwidth]{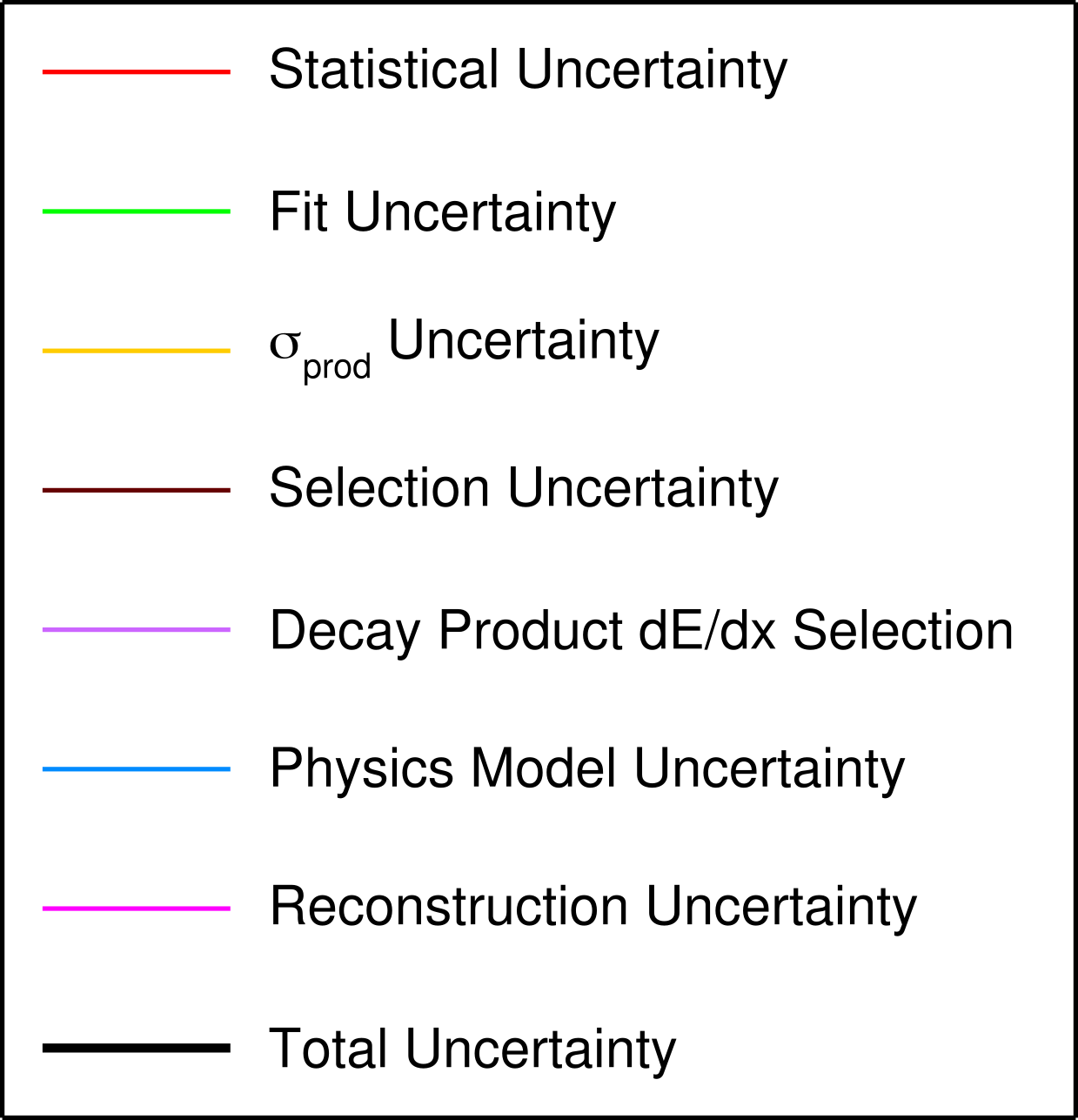}
  \includegraphics[width=0.33\textwidth]{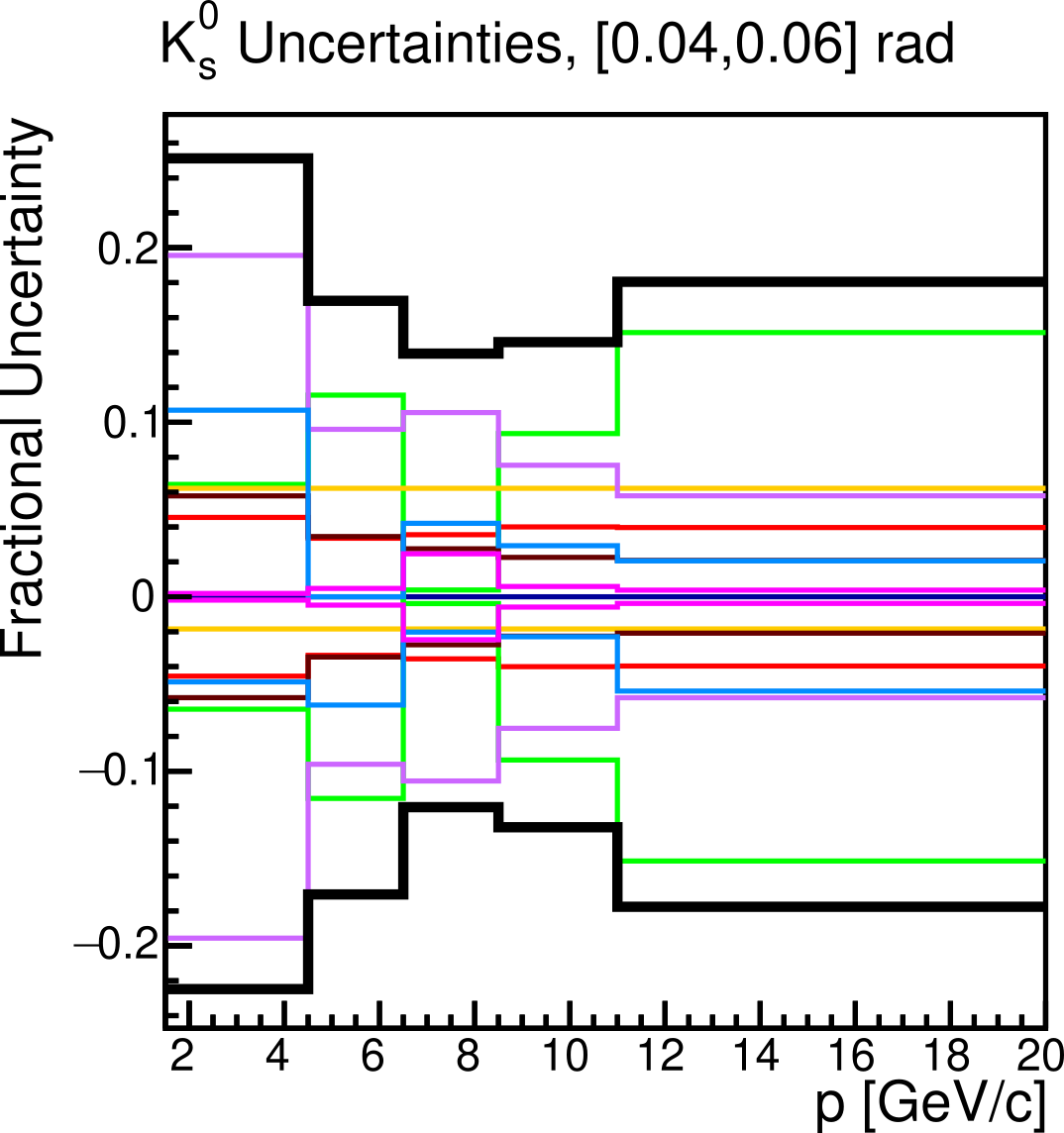}
  \includegraphics[width=0.33\textwidth]{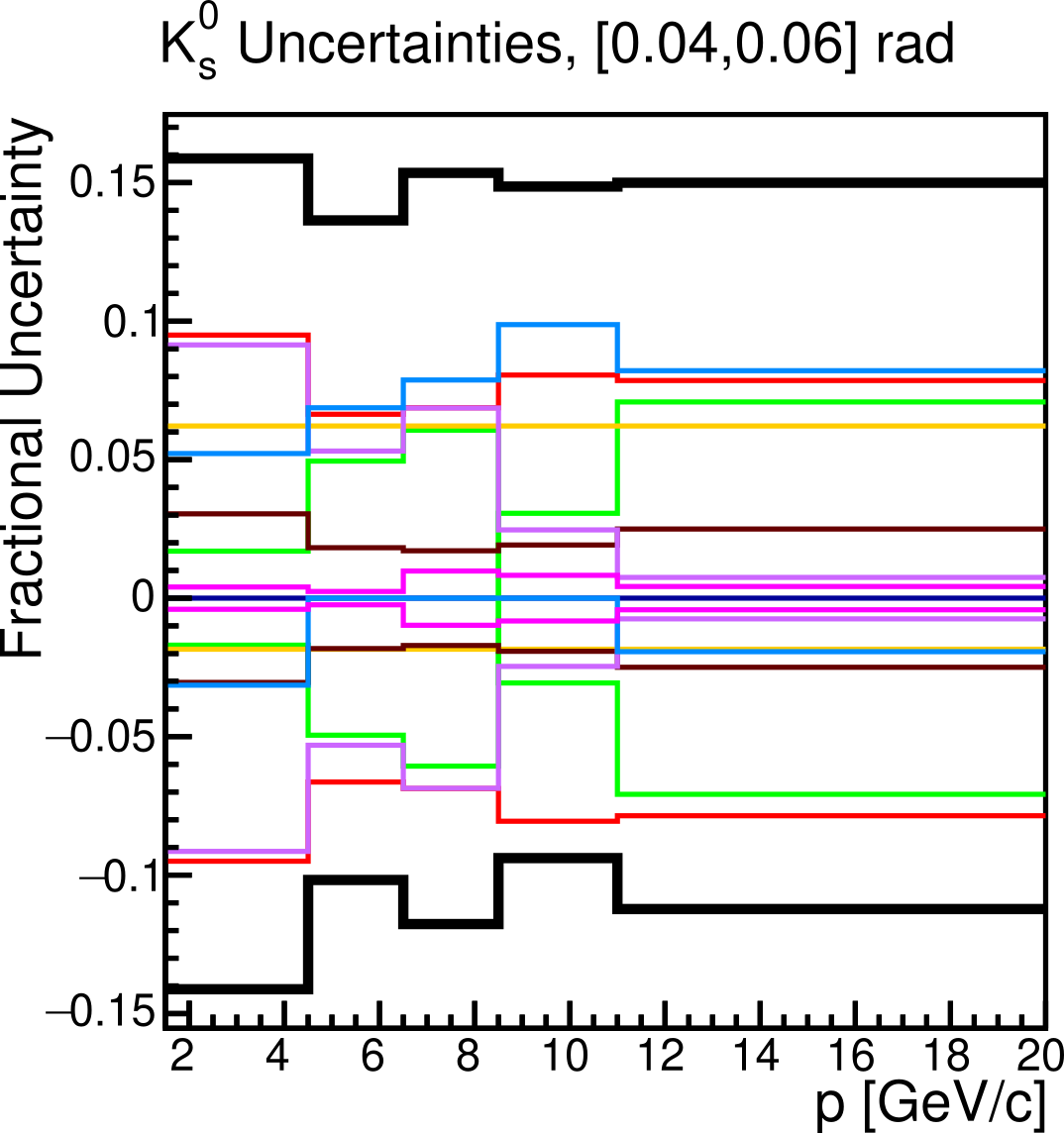}
\caption{Systematic uncertainty breakdown for 2016 and 2017 \kos analyses. One representative angular bin is shown.}\label{fig:systematicsK0S}
\end{figure*}

\begin{figure*}[!ht]
  \centering
  \includegraphics[width=0.3\textwidth]{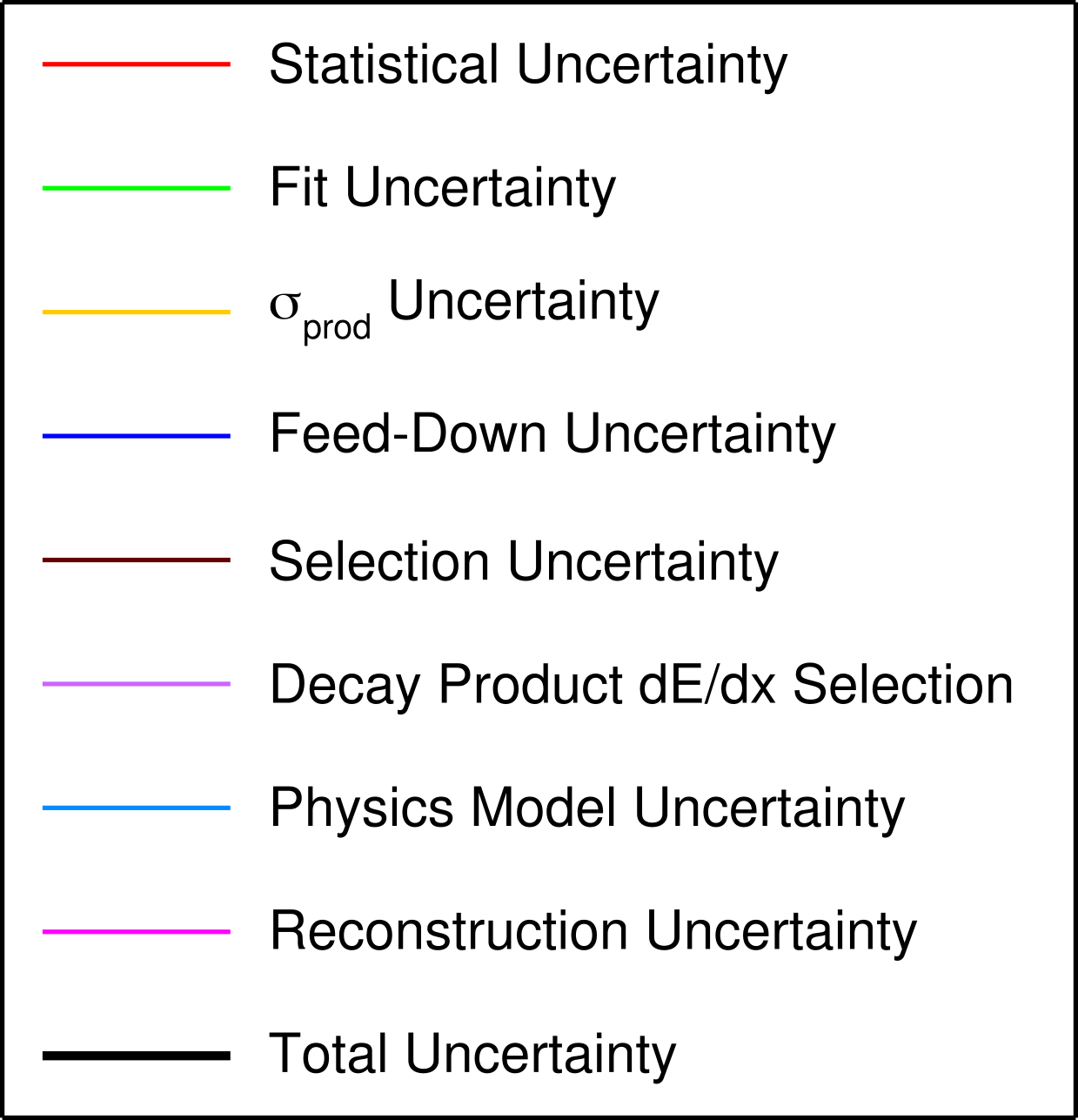}
  \includegraphics[width=0.33\textwidth]{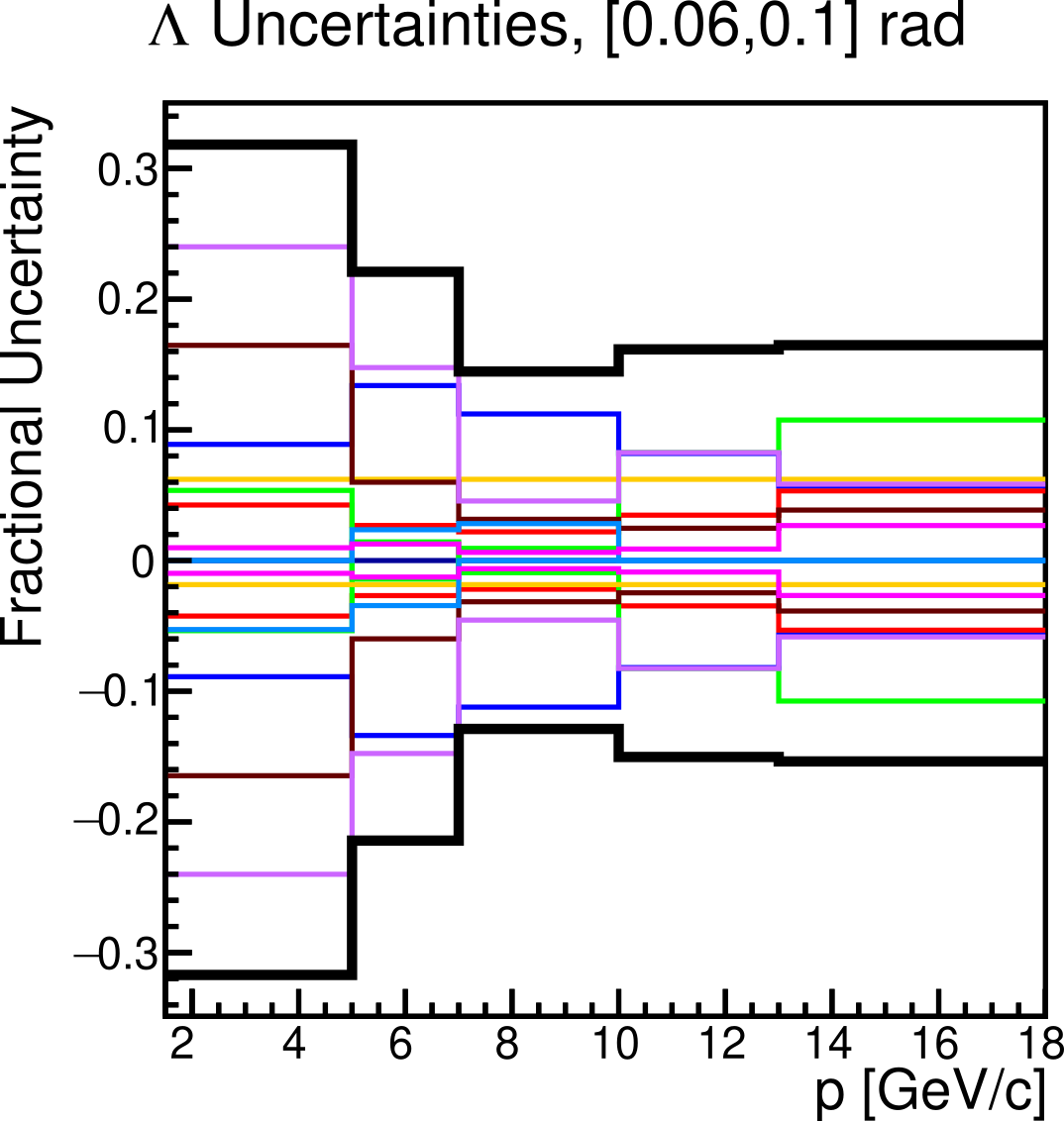}
  \includegraphics[width=0.33\textwidth]{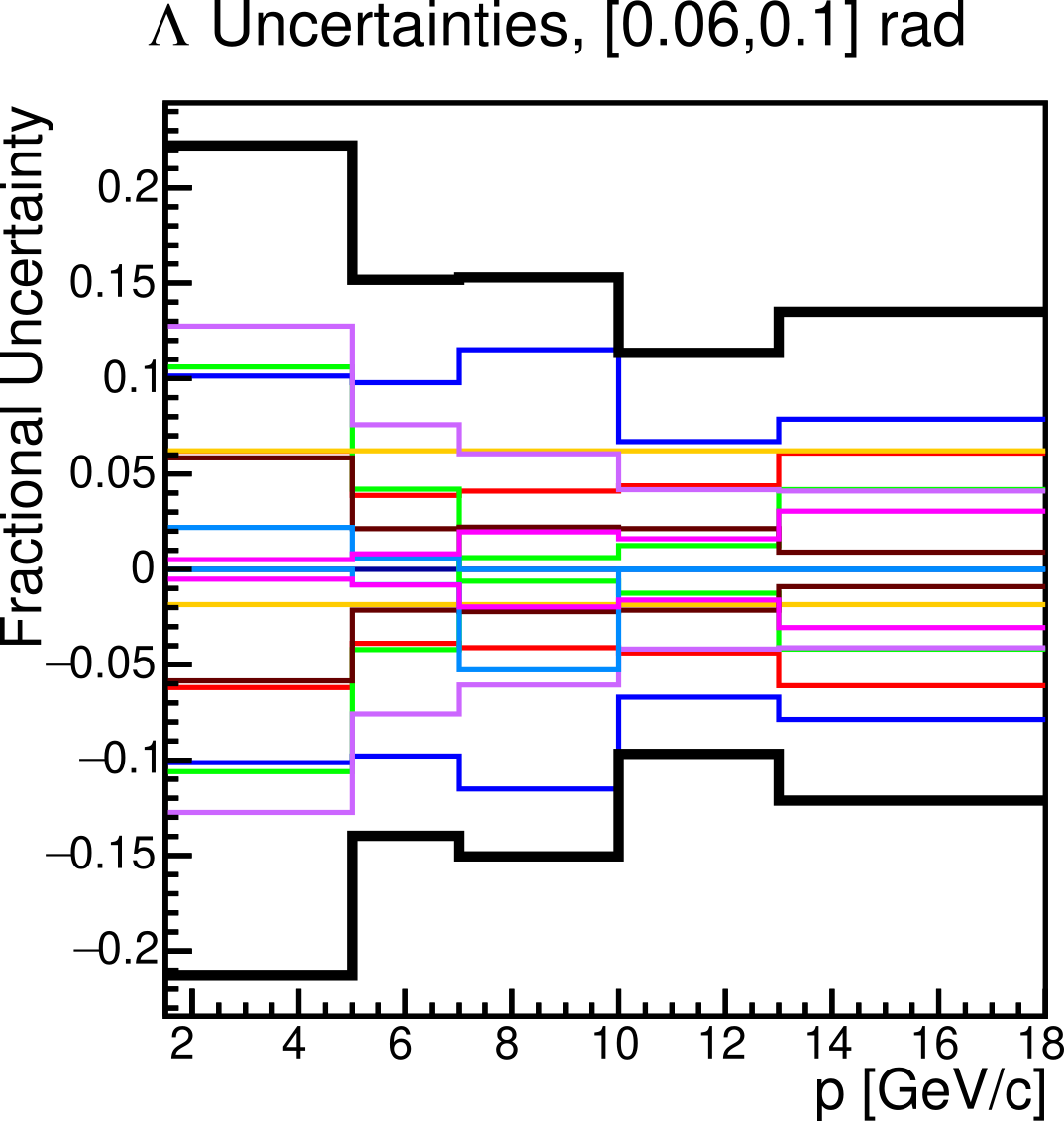}
\caption{Systematic uncertainty breakdown for 2016 and 2017 \lam analyses. One representative angular bin is shown.}\label{fig:systematicsLam}
\end{figure*}

\begin{figure*}[!ht]
  \centering
  \includegraphics[width=0.3\textwidth]{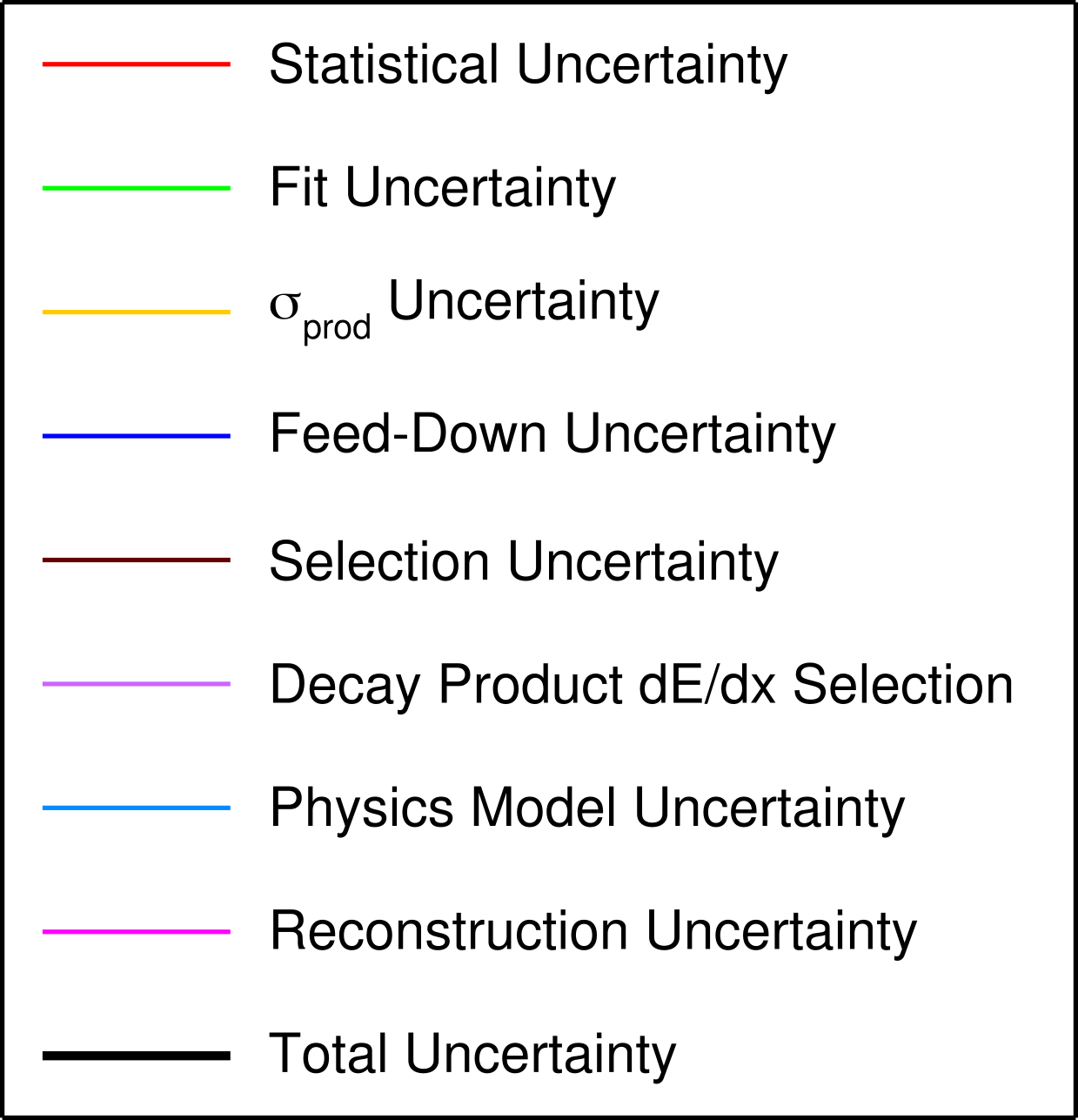}
  \includegraphics[width=0.33\textwidth]{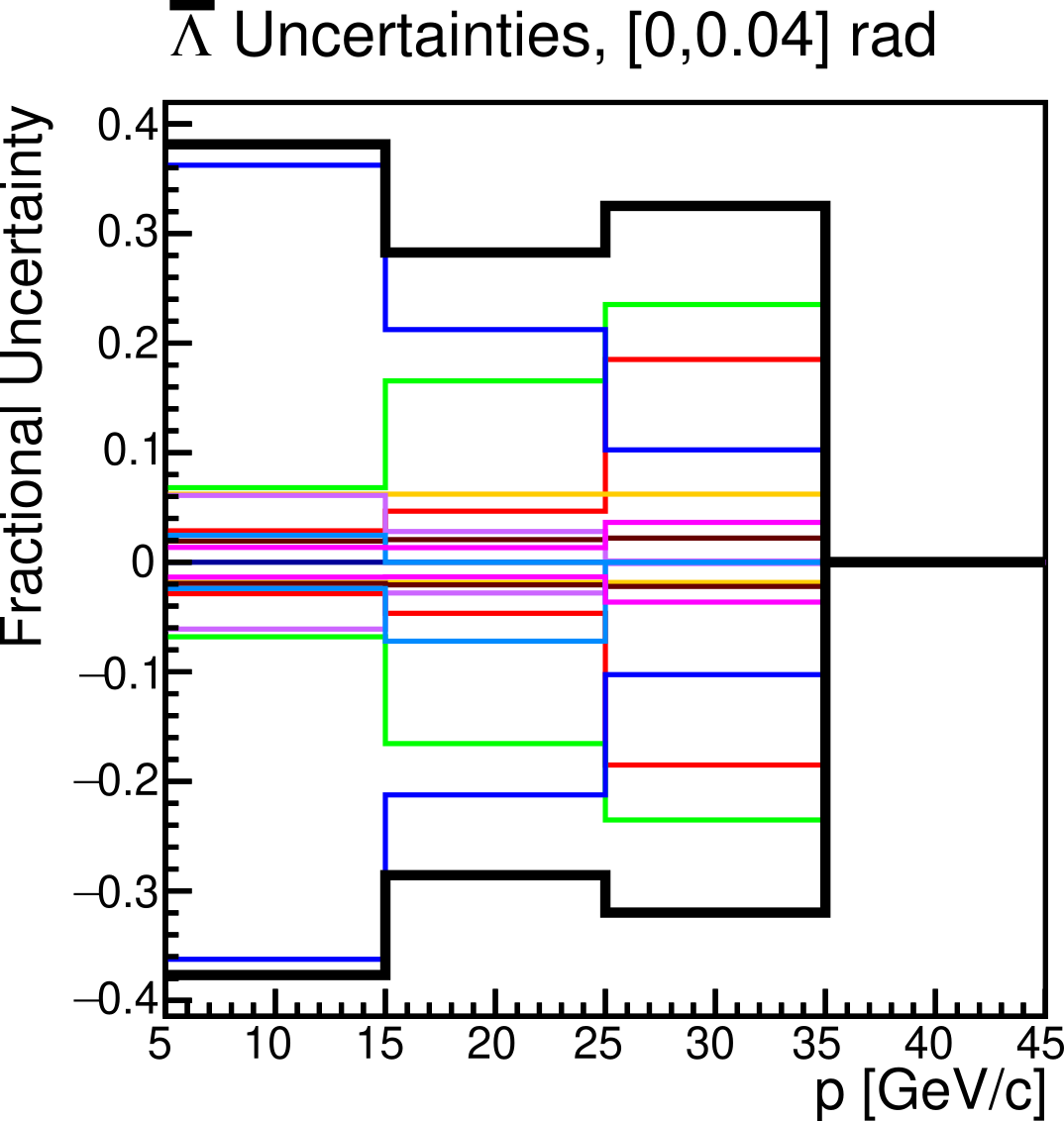}
  \includegraphics[width=0.33\textwidth]{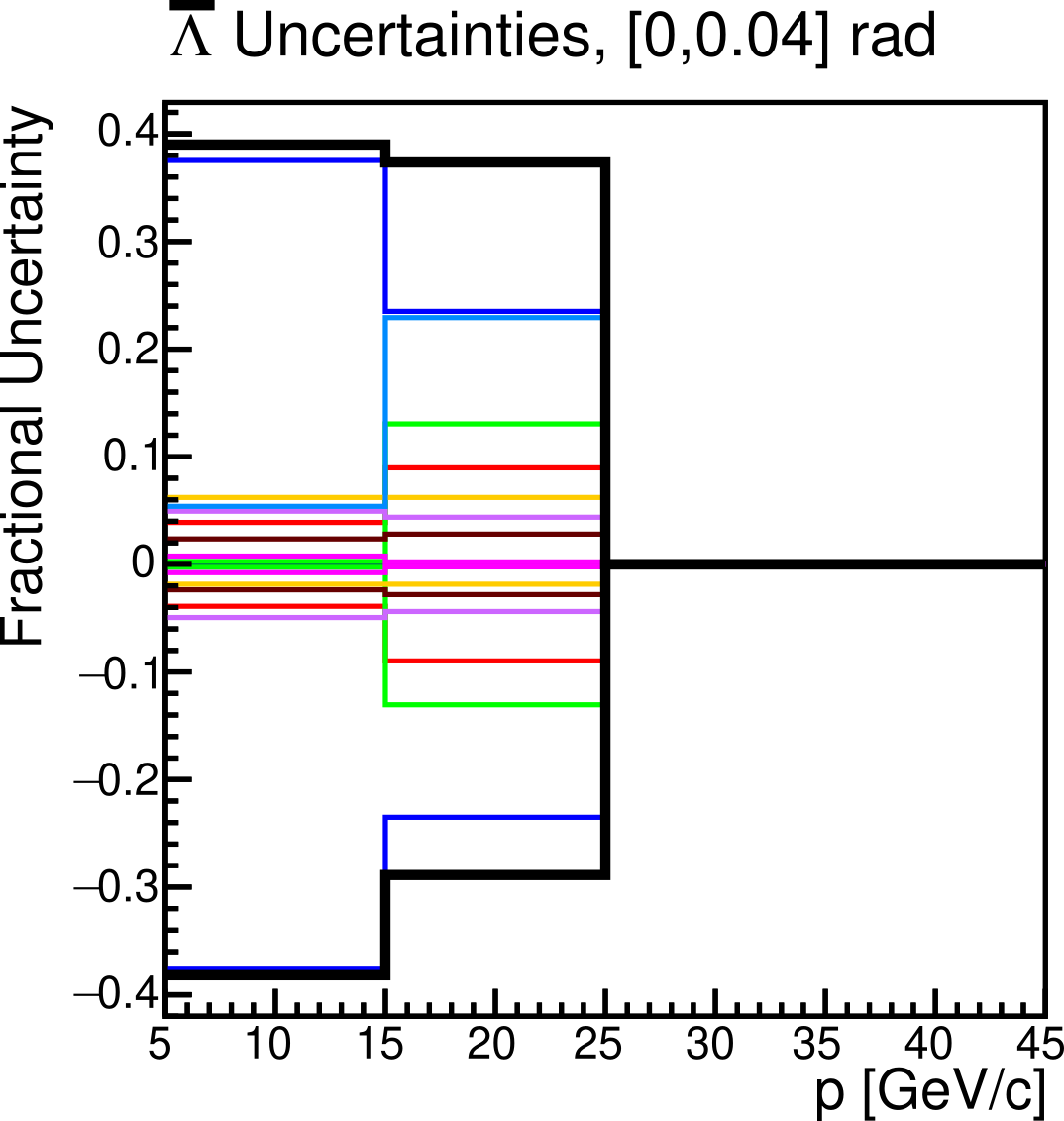}
\caption{Systematic uncertainty breakdown for 2016 and 2017 \alam analyses. One representative angular bin is shown. Regions with 0 total uncertainty correspond to bins where no multiplicity measurement was reported.}\label{fig:systematicsALam}
\end{figure*}

A breakdown of the neutral analysis systematic uncertainties can be seen in Figs.~\ref{fig:systematicsK0S}--~\ref{fig:systematicsALam}.

\subsection{Reconstruction}

Differences between true detector positions and those used in the Monte Carlo simulation affect final multiplicity measurements. Residual distributions describing track and point measurement mismatch were used to estimate potential detector misalignment. To estimate the reconstruction uncertainty, the detector central positions were displaced by varying amounts and the change in multiplicity was studied. VTPC1 and VTPC2 were simultaneously shifted by $\pm$200 \textmu{}m in opposite directions in the $x$-dimension, i.e. the central $x$-position of VTPC1 was shifted by +200 \textmu{}m while the central $x$-position of VTPC2 was shifted by -200 \textmu{}m, and vice-versa. The magnitudes of these shifts were motivated by the widths of the track residual distributions. Shifts in the $x$-dimension were found to have the most significant impact on momentum and track reconstruction, as $x$ is the bending plane in the magnetic field. The resulting multiplicity differences were added in quadrature to obtain the final reconstruction uncertainty.

\subsection{Selection}

Upon comparing track characteristics between reconstructed Monte Carlo and recorded data, a discrepancy was found in the average number of clusters per track. The simulated tracks contain 5 - 10 \% more clusters than tracks from data. This is likely due to unsimulated faulty front-end electronics channels and periodic detector noise. These two effects often lead to cluster loss, as the cluster structures become difficult to distinguish from background noise. In order to compensate for this effect, the Monte Carlo corrections were re-calculated after artificially reducing the number of clusters on the simulated track by 15\% for a conservative estimate. The resulting Monte Carlo corrections were used to re-calculate the multiplicity measurements, and the difference was taken as a systematic uncertainty.

\subsection{Physics Model}

The Monte Carlo correction factors are calculated using a given \GeantFour physics list. Varying the underlying physics list will lead to different correction factors. The central values for the Monte Carlo corrections were determined using the FTFP\_BERT physics list, which appears to be more consistent with NA61/SHINE data than other physics lists. Three other physics lists, FTF\_BIC, QGSP\_BERT, and QBBC were substituted in independent Monte Carlo samples, and the multiplicities were re-calculated with these correction factors. The difference from the nominal multiplicities was taken as a systematic uncertainty.

\subsection{Production Cross-Section Uncertainty}

The 120 \gevc proton-carbon production cross-section measurement was reported with a highly asymmetric systematic uncertainty~\cite{na61ProtonCrossSections}. The upper and lower uncertainty values were propagated through the multiplicity analysis in order to obtain the associated uncertainty on the multiplicity spectra. The result is a uniform fractional uncertainty on each measurement of (+5.8,-1.8)\%. This uncertainty can be significantly reduced in the future when a more precise measurement of the 120 \gevc proton-carbon quasi-elastic cross-section is made. 

\subsection{Momentum}

Uncertainty on the momentum reconstruction scale was estimated by studying the \kos invariant mass spectrum. An aggregate invariant mass sample was created by merging the kinematic analysis bins, and the \kos mass was fit for using a Breit-Wigner signal model and a 3rd-order polynomial background model. The fractional difference between the current accepted value for the \kos mass~\cite{PDG} and the aggregate fit mass was taken as a momentum uncertainty. The momenta of all tracks were then shifted by this amount and the resulting change in multiplicities was taken as a systematic uncertainty. For the 2016 analysis the measured mass shift was $\Delta m = -0.1$ \mev (-0.02\%) and for the 2017 analysis the measured mass shift was $\Delta m = 1.1$ \mev (0.22\%). This uncertainty source was significantly smaller than the other systematic uncertainties and thus was not included in the uncertainty evaluation. 

\subsection{Feed-down}

The feed-down uncertainty for the neutral-hadron analysis is based solely on Monte Carlo feed-down estimates. Weak decays of $\Xi$ and $\Omega$ baryons producing \lam and \alam were considered. Production rates of these baryons vary among physics lists up to 50\%. In order to estimate the uncertainty associated with these decays, the number of feed-down tracks was varied by $\pm$ 50\% and the feed-down correction factor was re-calculated. The resulting changes in multiplicities are taken as a systematic uncertainty. The resulting uncertainties are (5--10)\% for \lam and (5--20)\% for \alam.

\subsection{$dE/dx$ Selection}

An uncertainty associated with $dE/dx$ selection of decay products was calculated by relaxing the cut by 5\%. The data and Monte Carlo corrections were then re-processed, and the invariant mass fits to the varied data samples were performed in each kinematic bin. The resulting yields were used to calculate new multiplicities, and the changes in multiplicity were taken as a systematic uncertainty.

\subsection{Invariant Mass Fit}

An uncertainty associated with the invariant mass signal fit was estimated using four \GeantFour Monte Carlo physics lists (FTFP\_BERT, QGSP\_BERT, QBBC, FTF\_BIC). Invariant mass fits for each kinematic bin were performed and the number of fit signal tracks was compared to the true number of signal tracks. The fractional differences were averaged to estimate the fit uncertainty, and the average difference was taken as a systematic uncertainty.

\section{Combined Multiplicity Measurements}\label{sec:combinedMultiplicities}

In regions of phase space where detector acceptance overlapped in 2016 and 2017, multiplicity measurements are combined. The measurements must be weighted by the square of the total uncertainty specific to each analysis, referred to here as the uncorrelated uncertainty. This uncertainty includes statistical, reconstruction, selection, momentum, and fit uncertainties. Correlated uncertainty, consisting of feed-down, production cross-section, and physics model uncertainties, applies to both analyses and are not included in measurement weights during combination.

For the combined multiplicity measurement, a simple weighted mean is calculated using the uncorrelated uncertainty, which consists of symmetric uncertainties only:

\begin{equation}
m_\textrm{combined} = 
\frac{\frac{m_1}{\sigma_1^2} + \frac{m_2}{\sigma_2^2}}{\frac{1}{\sigma_1^2} + \frac{1}{\sigma_2^2}},
\end{equation}

where $m_1$ and $\sigma_1$ are the multiplicity measurement and uncorrelated uncertainty from the 2016 analysis and $m_2$ and $\sigma_2$ are the corresponding values from the 2017 analysis.

A reduced $\chi^2$ value was calculated for each analysis reflecting the compatibility of the 2016 and 2017 multiplicity measurements. The $\chi^2$ values, numbers of degrees of freedom, and corresponding p-values are presented in Table~\ref{tab:chi2}. In general, the measurements agree well, with a reduced $\chi^2$ near 1. These values were calculated using the differences in each measurement and the measurement covariance matrices representing the systematic uncertainties specific to each analysis. The resulting $\chi^2$, $NDF$, and $p$-values are quoted in Table~\ref{tab:chi2}. The covariance matrices are available at~\cite{pC120EDMS}. 

\begin{table*}[htbp]
\centering
\begin{tabular}{cccccccc}
Neutral Hadron Species & $\chi^2$ & NDF & $p$-value \\
\hline
\kos  & 38.5 & 34 & 0.27 \\
\lam  & 13.6 & 26 & 0.97 \\
\alam & 3.8  &  5 & 0.58 \\

\end{tabular}
\caption[Measurement $\chi^2$]{$\chi^2$ values corresponding to the combination of the 2016 and 2017 multiplicity measurements.}
\label{tab:chi2}
\end{table*}

\begin{figure*}[!ht]
  \centering
  \includegraphics[width=\textwidth]{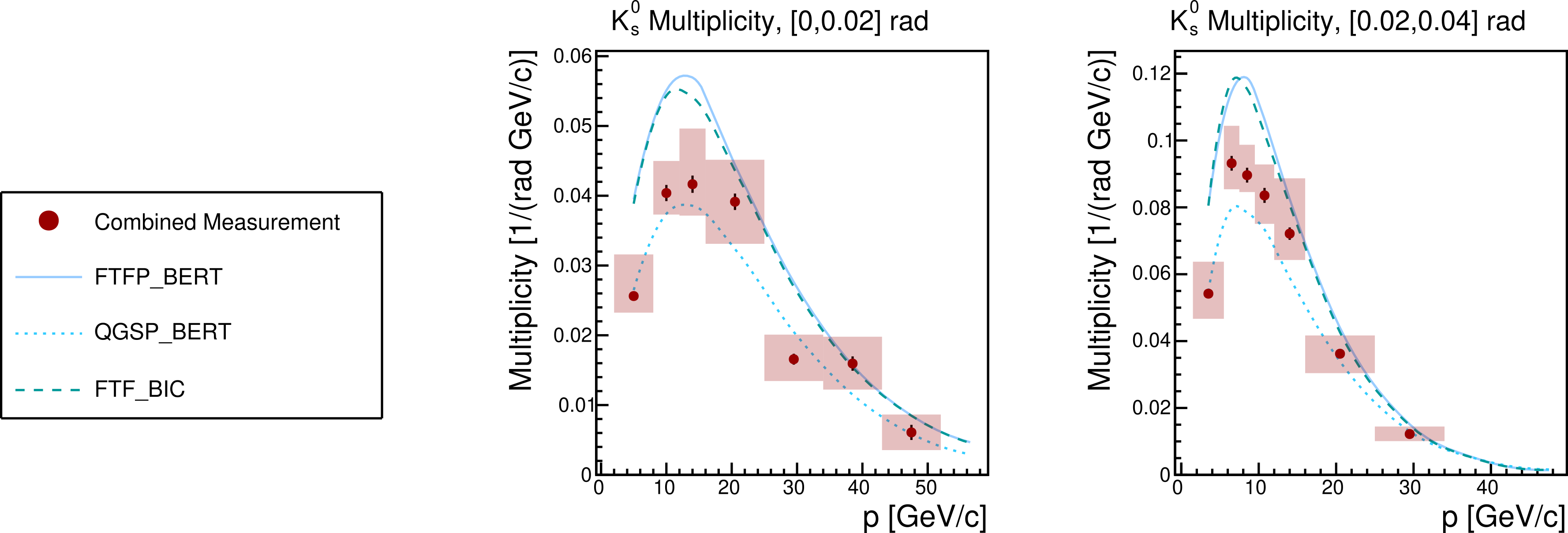}
\caption{Combined multiplicity measurements for \kos analysis. Error bars denote statistical uncertainty, and total systematic uncertainty is shown as a red band. Results are compared to three \GeantFour physics lists. Two representative angular bins are shown.}\label{fig:combinedResultK0S}
\end{figure*}

\begin{figure*}[!ht]
  \centering
  \includegraphics[width=\textwidth]{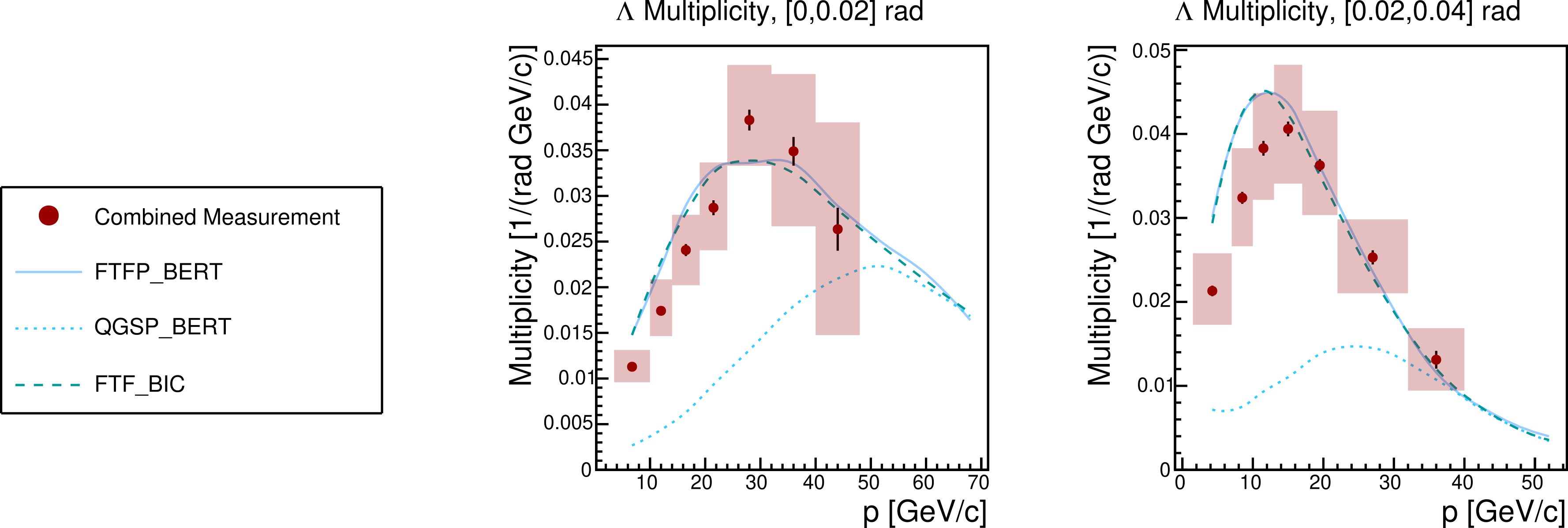}
\caption{Combined multiplicity measurements for \lam analysis. Error bars denote statistical uncertainty, and total systematic uncertainty is shown as a red band. Results are compared to three \GeantFour physics lists. Two representative angular bins are shown.}\label{fig:combinedResultLam}
\end{figure*}

\begin{figure*}[!ht]
  \centering
  \includegraphics[width=\textwidth]{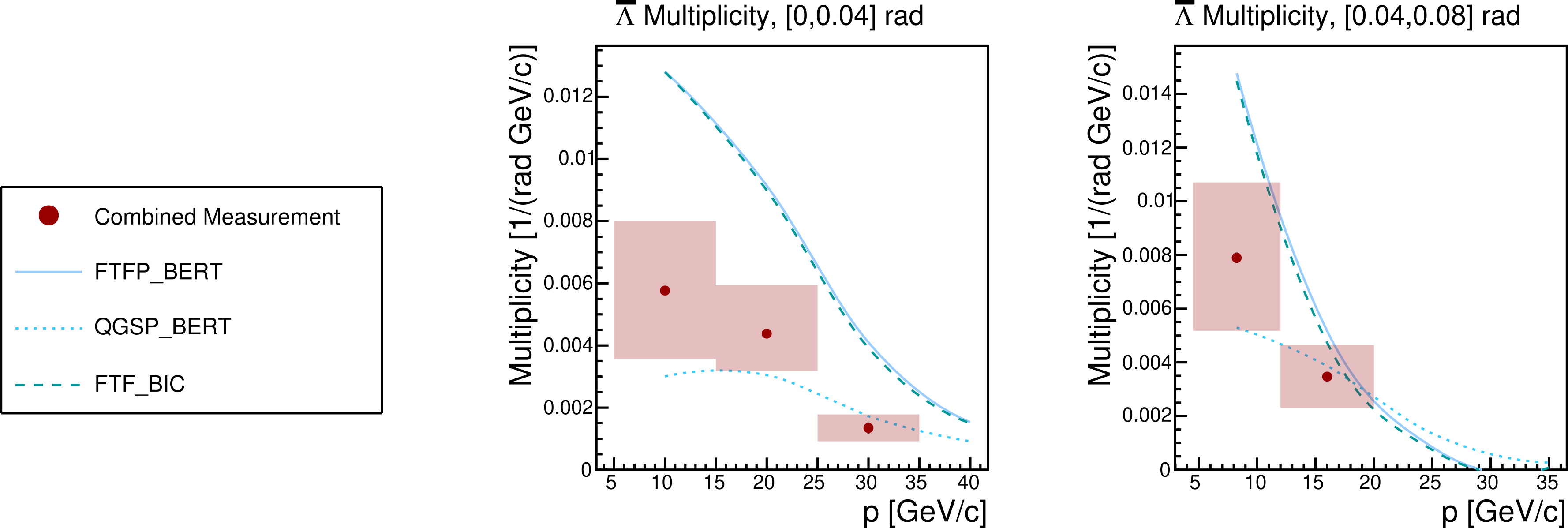}
\caption{Combined multiplicity measurements for \alam analysis. Error bars denote statistical uncertainty, and total systematic uncertainty is shown as a red band. Results are compared to three \GeantFour physics lists. Two representative angular bins are shown.}\label{fig:combinedResultALam}
\end{figure*}

Combined multiplicity results for the neutral-hadron analysis can be seen in Figs.~\ref{fig:combinedResultK0S}--~\ref{fig:combinedResultALam}. 

\subsection{Combined Systematic Uncertainties}

The total systematic uncertainties on the combined multiplicities reflect both the uncorrelated uncertainties unique to each analysis and the correlated uncertainties that apply to both analyses. Fractional uncorrelated uncertainties are added in quadrature and applied to the combined multiplicity value. Fractional correlated uncertainties are treated differently, as they should not simply be added in quadrature. For each correlated uncertainty in each analysis bin, the fractional uncertainties were compared between the 2016 and 2017 analyses. The larger of the two was taken as the total contribution to the total uncertainty. The final values for the uncorrelated uncertainty and each correlated uncertainty were added in quadrature to obtain the total systematic uncertainty.

Uncertainties considered to be uncorrelated between the two analyses are the statistical uncertainty, invariant mass fit uncertainty, decay product $dE/dx$ selection uncertainty, reconstruction uncertainty, and $\vo$ selection uncertainty. These uncertainties are considered uncorrelated due to significant differences in detector configuration for the 2016 and 2017 data sets, which results in different phase space occupancy. Uncertainties considered to be correlated between the two analyses are feed-down uncertainty, production cross-section uncertainty, and physics model uncertainty. 

\begin{figure*}[!ht]
  \centering
  \includegraphics[width=\textwidth]{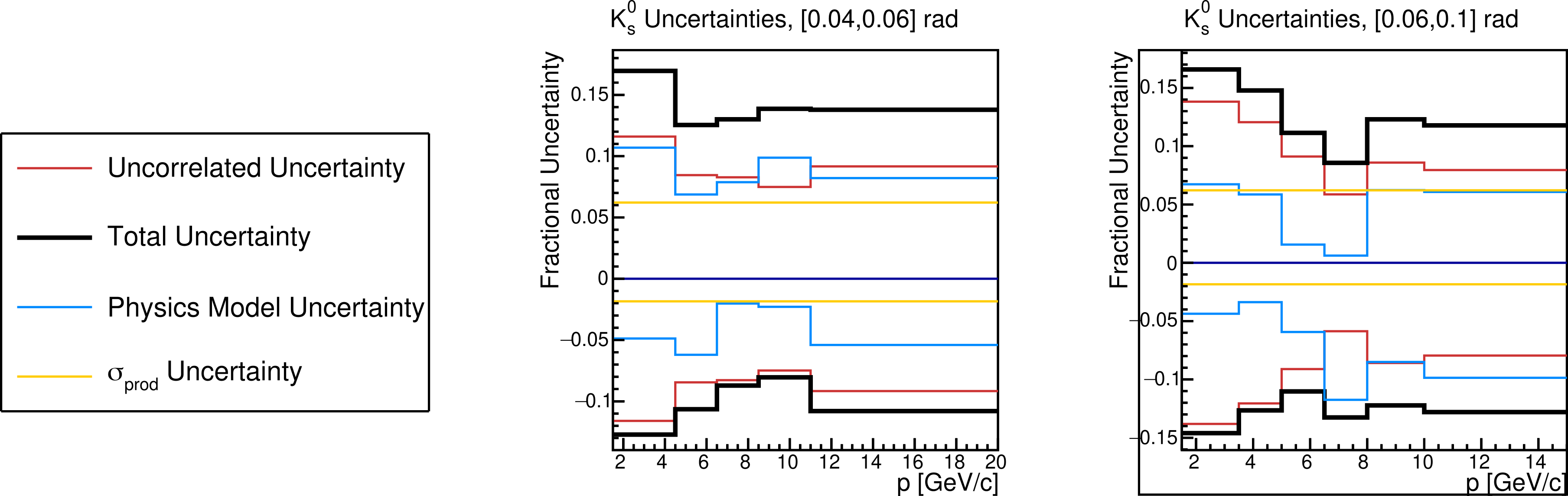}
\caption{Systematic uncertainty breakdown for the combined \kos analysis. Two representative angular bins are shown.}\label{fig:combinedUncertaintiesK0S}
\end{figure*}

\begin{figure*}[!ht]
  \centering
  \includegraphics[width=\textwidth]{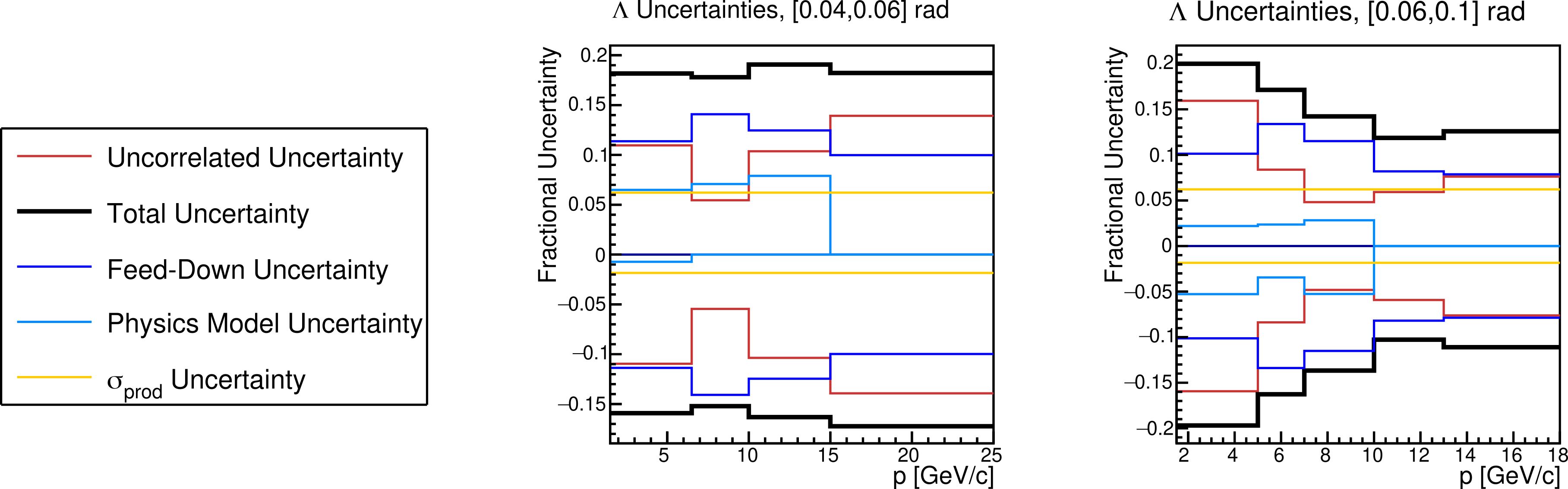}
\caption{Systematic uncertainty breakdown for the combined \lam analysis. Two representative angular bins are shown.}\label{fig:combinedUncertaintiesLam}
\end{figure*}

\begin{figure*}[!ht]
  \centering
  \includegraphics[width=\textwidth]{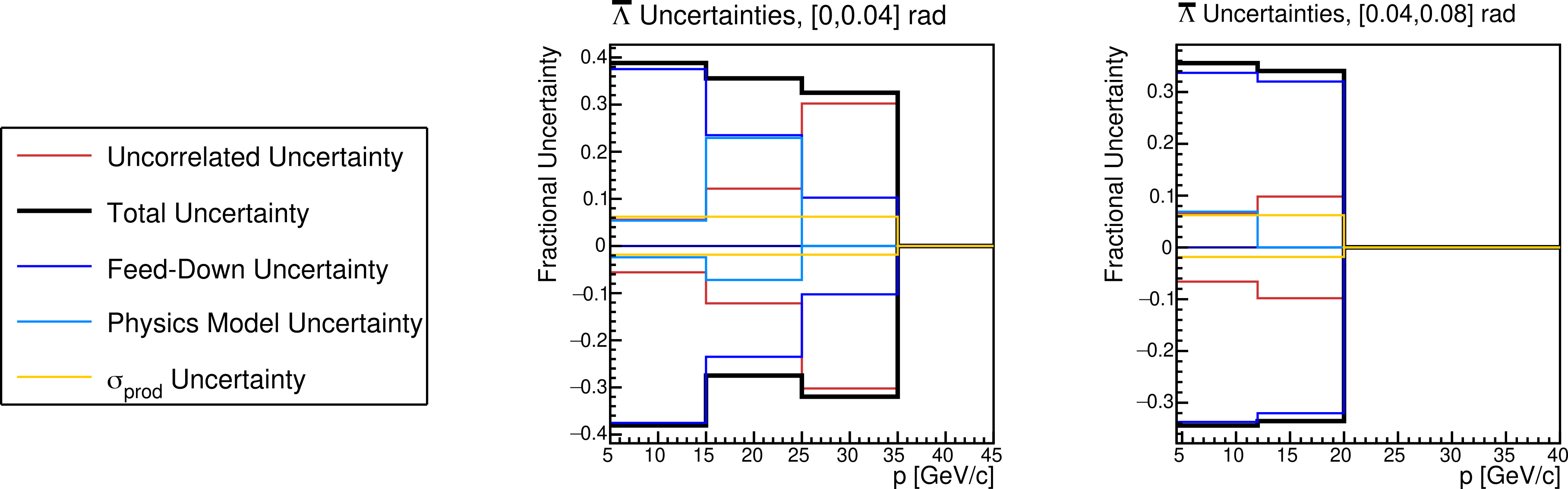}
\caption{Systematic uncertainty breakdown for the combined \alam analysis. Regions with 0 total uncertainty correspond to bins where no multiplicity measurement was reported. }\label{fig:combinedUncertaintiesALam}
\end{figure*}

A breakdown of the neutral analysis combined systematic uncertainties can be seen in Figs.~\ref{fig:combinedUncertaintiesK0S}--~\ref{fig:combinedUncertaintiesALam}.

\section{Summary}\label{sec:Summary}

Measurements of neutral-hadron production in 120 \gevc proton-carbon interactions were presented. The results are the combination of two complementary data sets recorded with significantly different detector configurations. Agreement in overlapping regions of phase space strengthens the results. Significant discrepancies between the measurements and popular Monte Carlo simulation physics lists were highlighted. In particular, \lam production in the \GeantFour QGSP\_BERT physics list shows significant discrepancy with these measurements, and \alam production in all \GeantFour physics lists show significant discrepancy with measured \alam multiplicities. 

Dominant systematic uncertainties in the neutral-hadron analysis originate from invariant mass spectrum fits. These uncertainties could be reduced by increasing the collected number of events, or by reducing the background in the invariant mass spectra. $dE/dx$ decay product selection also incurs a significant systematic uncertainty. This uncertainty could be reduced by improving the quality of $dE/dx$ calibration.

Numerical results of the multiplicity measurements of \kos, \lam and \alam are summarized in CERN EDMS~\cite{pC120EDMS} along with statistical, systematic and total uncertainties for each kinematic bin. Covariance matrices for each analysis are included.

The results presented in this publication can be used to improve the accuracy of neutrino beam content estimation in existing and future experiments in which the neutrino beam is created using the 120 \gevc proton-carbon interaction. In addition, these multiplicity measurements will be used in a forthcoming publication to constrain feed-down contributions to charged-hadron multiplicity measurements, which will result in significant reduction of systematic uncertainties associated with feed-down decays.

  \section*{Acknowledgments}


We would like to thank the CERN EP, BE, HSE and EN Departments for the
strong support of NA61/SHINE.

This work was supported by
the Hungarian Scientific Research Fund (grant NKFIH 138136\slash138152),
the Polish Ministry of Science and Higher Education
(DIR\slash WK\slash\-2016\slash 2017\slash\-10-1, WUT ID-UB), the National Science Centre Poland (grants
2014\slash 14\slash E\slash ST2\slash 00018, 
2016\slash 21\slash D\slash ST2\slash 01983, 
2017\slash 25\slash N\slash ST2\slash 02575, 
2018\slash 29\slash N\slash ST2\slash 02595, 
2018\slash 30\slash A\slash ST2\slash 00226, 
2018\slash 31\slash G\slash ST2\slash 03910, 
2019\slash 33\slash B\slash ST9\slash 03059, 
2020\slash 39\slash O\slash ST2\slash 00277), 
the Norwegian Financial Mechanism 2014--2021 (grant 2019\slash 34\slash H\slash ST2\slash 00585),
the Polish Minister of Education and Science (contract No. 2021\slash WK\slash 10),
the Russian Science Foundation (grant 17-72-20045),
the Russian Academy of Science and the
Russian Foundation for Basic Research (grants 08-02-00018, 09-02-00664
and 12-02-91503-CERN),
the Russian Foundation for Basic Research (RFBR) funding within the research project no. 18-02-40086,
the Ministry of Science and Higher Education of the Russian Federation, Project "Fundamental properties of elementary particles and cosmology" No 0723-2020-0041,
the European Union's Horizon 2020 research and innovation programme under grant agreement No. 871072,
the Ministry of Education, Culture, Sports,
Science and Tech\-no\-lo\-gy, Japan, Grant-in-Aid for Sci\-en\-ti\-fic
Research (grants 18071005, 19034011, 19740162, 20740160 and 20039012),
the German Research Foundation DFG (grants GA\,1480\slash8-1 and project 426579465),
the Bulgarian Ministry of Education and Science within the National
Roadmap for Research Infrastructures 2020--2027, contract No. D01-374/18.12.2020,
Ministry of Education
and Science of the Republic of Serbia (grant OI171002), Swiss
Nationalfonds Foundation (grant 200020\-117913/1), ETH Research Grant
TH-01\,07-3 and the Fermi National Accelerator Laboratory (Fermilab), a U.S. Department of Energy, Office of Science, HEP User Facility managed by Fermi Research Alliance, LLC (FRA), acting under Contract No. DE-AC02-07CH11359 and the IN2P3-CNRS (France).\\

The data used in this paper were collected before February 2022.


\bibliography{na61Preprint}

\clearpage

\end{document}